\documentclass[12pt,journal,draftcls,a4paper,onecolumn]{IEEEtran}

\usepackage{dsfont}
\usepackage{amsmath}
\usepackage{amssymb}
\usepackage{amsfonts}
\usepackage{graphicx}
\usepackage{amsmath}
\usepackage[all,poly]{xy}
\usepackage{multirow}
\usepackage{algorithm}
\usepackage{algorithmic}
\usepackage{color}
\usepackage[noadjust]{cite}

\newcommand{\figwidth}{\columnwidth}

\newcommand{\figextension}{png}
\graphicspath{{figures/}}

\def\bss{{\boldsymbol{s}}}

\def\bsv{{\boldsymbol{v}}}

\def\bsB{{\boldsymbol{B}}}

\def\bsG{{\boldsymbol{G}}}

\def\bsN{{\boldsymbol{N}}}

\def\bfa{{\mathbf{a}}}

\def\bfs{{\mathbf{s}}}

\def\bfD{{\mathbf{D}}}

\def\bfU{{\mathbf{U}}}
\def\bfV{{\mathbf{V}}}

\def\bfX{{\mathbf{X}}}

\def\calU{{\mathcal{U}}}

\def\calS{{\mathcal{S}}}

\def\calA{{\mathcal{A}}}
\def\calB{{\mathcal{B}}}

\def\calG{{\mathcal{G}}}

\def\calI{{\mathcal{I}}}

\def\calN{{\mathcal{N}}}
\def\calO{{\mathcal{O}}}

\def\calS{{\mathcal{S}}}

\def\calU{{\mathcal{U}}}

\def\calX{{\mathcal{X}}}

\newcommand{\SNR}{\mathrm{SNR}}

\newcommand{\Vobs}[1]{\boldsymbol{x}(#1)}
\newcommand{\MATobs}{\boldsymbol{X}}
\newcommand{\obs}[2]{x_{#1}(#2)}

\newcommand{\Vsou}[1]{\boldsymbol{s}(#1)}
\newcommand{\VsouP}[2]{\boldsymbol{s}_{#1}(#2)}
\newcommand{\MATsou}{\boldsymbol{S}}
\newcommand{\sou}[2]{s_{#1}(#2)}

\newcommand{\souvar}[1]{a_{#1}^2}
\newcommand{\Vsouvar}{\bfa^2}
\newcommand{\souproba}[1]{\lambda_{#1}}
\newcommand{\Vsouproba}{\boldsymbol{\lambda}}

\newcommand{\Vindic}[1]{\boldsymbol{q}(#1)}
\newcommand{\VindicP}[2]{\boldsymbol{q}_{#2}(#1)}

\newcommand{\MATindic}{\boldsymbol{Q}}
\newcommand{\indic}[2]{q_{#1}(#2)}

\newcommand{\dimobs}{M}

\newcommand{\nbobs}{T}
\newcommand{\noobs}{t}

\newcommand{\nbsou}{N}
\newcommand{\nosou}{n}

\newcommand{\MATaxe}{\boldsymbol{\Psi}}
\newcommand{\Vaxe}[1]{\boldsymbol{\psi}_{#1}}
\newcommand{\axe}[2]{\psi_{#1,#2}}

\newcommand{\Vnoise}[1]{{\boldsymbol{n}}({#1})}

\newcommand{\MATnoise}{{\boldsymbol{N}}}

\newcommand{\noisevar}{{\sigma}^2}

\newcommand{\R}{\mathds{R}}

\newcommand{\paramvect}{\boldsymbol{\theta}}
\newcommand{\hypervect}{\boldsymbol{\phi}}

\newcommand{\sample}[2]{\widetilde{#1}^{(#2)}}
\newcommand{\samplenoisevar}[1]{{\widetilde{\sigma}}^{2(#1)}}
\newcommand{\sampleVsouvar}[1]{{\widetilde{\bfa}}^{2(#1)}}

\newcommand{\inv}{^{-1}}

\newcommand{\transp}{^T}

\newcommand{\proba}[1]{\mathrm{P}\left[#1\right]}

\newcommand{\norm}[1]{\left\|#1\right\|}

\newcommand{\Steifel}[1]{\calS_{#1}}

\newcommand{\vol}[1]{\mathrm{vol}\left(#1\right)}

\newcommand{\dirac}[1]{\delta\left(#1\right)}

\newcommand{\Vzero}[1]{\boldsymbol{0}_{#1}}
\newcommand{\Id}[1]{\textbf{I}_{#1}}
\newcommand{\Indicfun}[2]{\textbf{1}_{#1}\left(#2\right)}

\newcounter{algo}
\renewcommand{\thealgo}{\arabic{algo}}

\title{Bayesian Orthogonal Component \\Analysis for Sparse Representation}

\author{Nicolas Dobigeon and Jean-Yves Tourneret
\\
\normalsize University of Toulouse, IRIT/INP-ENSEEIHT/T\'eSA\\
2 rue Camichel, 31071 Toulouse, France. \\
\small\texttt{\{Nicolas.Dobigeon,Jean-Yves.Tourneret\}@enseeiht.fr}}

\begin{document}

\maketitle

\begin{abstract}
This paper addresses the problem of identifying a lower dimensional
space where observed data can be sparsely represented. This
under-complete dictionary learning task can be formulated as a blind
separation problem of sparse sources linearly mixed with an unknown
orthogonal mixing matrix. This issue is formulated in a Bayesian
framework. First, the unknown sparse sources are modeled as
Bernoulli-Gaussian processes. To promote sparsity, a weighted
mixture of an atom at zero and a Gaussian distribution is proposed
as prior distribution for the unobserved sources. A non-informative
prior distribution defined on an appropriate Stiefel manifold is
elected for the mixing matrix. The Bayesian inference on the unknown
parameters is conducted using a Markov chain Monte Carlo (MCMC)
method. A partially collapsed Gibbs sampler is designed to generate
samples asymptotically distributed according to the joint posterior
distribution of the unknown model parameters and hyperparameters.
These samples are then used to approximate the joint maximum \emph{a
posteriori} estimator of the sources and mixing matrix. Simulations
conducted on synthetic data are reported to illustrate the
performance of the method for recovering sparse representations. An
application to sparse coding on under-complete dictionary is finally
investigated.
\end{abstract}


\hyphenation{hie-rar-chi-cal}
\newpage
\section{Introduction}

\PARstart{I}{n} recent years, sparse representations have motivated much research
in the signal processing community. This issue consists of
identifying a sparse decomposition of a signal on a given
dictionary. Among the main motivations, such representations have
been demonstrated to be an efficient alternative for regularizing
ill-posed inverse problems \cite{Gorodnitsky1997}. More recently,
compressive sensing has extensively benefited from sparsity to
reconstruct a signal from a few projections
\cite{Donoho2006,Baraniuk2007}. Signal reconstruction under hard
sparse constraints can be mainly formulated as an optimization
problem of a $\ell_0$-penalized quadratic criterion, whose numerical
resolution is unfortunately an NP-complete problem. Several greedy
algorithms have been proposed to approximate the signal
reconstruction solutions, such as the well-known matching pursuit
(MP) \cite{Mallat1993} and orthogonal matching pursuit (OMP)
\cite{Tropp2004} algorithms. However, under appropriate sufficient
conditions, replacing the $\ell_0$-norm by the $\ell_1$-norm in the
penalization term can lead to the same unique solution
\cite{Donoho2001}. Therefore, exploiting these interesting
sparseness properties, extensive works have been devoted on
$\ell_1$-constrained estimation problems for sparse representation
(see for example \cite{Alliney1994} and \cite{Gribonval2003}).

In all the above works, the (generally over-complete) dictionary on
which the signal is sparsely decomposed is assumed to be \emph{a
priori} known. The joint estimation of the atoms of the dictionary
and the corresponding sparse representation is a much more
challenging task. In \cite{Aharon2006}, Aharon \emph{et al.}
introduced an MP-based iterative method for designing over-complete
dictionaries. Of course, the over-completeness allows redundancy in
the atom decomposition. We address here the problem of recovering a
sparse data representation in a lower-dimensional space defined by
an under-complete orthogonal dictionary. Some up-to-date research
activities conducted in the signal processing and machine learning
communities have been focusing on this still open problem.
Specifically, Mishali and Eldar have introduced in
\cite{Mishali2009icassp} an alternating minimization procedure to
solve a complete sparse representation problem when the sparsity
level is assumed to be known. In \cite{Zou2006},
\cite{Aspremont2007}, and more recently in \cite{Aspremont2008} the
decomposition of a covariance matrix into sparse factors has been
formulated as a regression problem with sparsity constraints. More
generally, these matrix factorization strategies under some
particular constraints, e.g., non-negativity, orthogonality and
sparsity have demonstrated great interest for many different
applications. These applications include representation of natural
images \cite{Hoyer2004} and gene expression data analysis
\cite{Brunet2004}.

In this paper, the under-complete dictionary learning task is
formulated as a blind source separation problem with sparsity
constraints. Many applications have encouraged research on sparse
signal and image deconvolution. These applications include astronomy
\cite{Bourguignon2005ssp}, geophysics \cite{Cheng1996}, audio signal
decomposition \cite{Fevotte2006} and, more recently, molecular
imaging \cite{Dobigeon2009ip}. In the present work, we propose a
hierarchical Bayesian model for blind separation of sparse sources
linearly mixed by an orthogonal matrix\footnote{In the following,
the mixing matrix is said orthogonal although it is not a square
matrix. This abuse of language will mean that its columns, i.e., the
dictionary atoms, are orthogonal.}. This model is based on the
choice of pertinent prior distributions for unknown parameters and
hyperparameters. Following the works of Kormylo and Mendel
\cite{Kormylo1982}, the unknown sources are assumed to be
Bernoulli-Gaussian (BG) processes. Therefore, the source prior is
composed of a weighted mixture of a standard Gaussian distribution
and a mass at zero. Note that this distribution has been widely
advocated to solve reconstruction problems in a Bayesian framework
(see \cite{Idier1990,Lavielle1993,Doucet1997,Fevotte2008} among
others). However, estimating hyperparameters involved in such prior
mixture is a critical issue that drastically impacts the estimation
performance. As an example, the empirical Bayes (EB) and Stein
unbiased risk (SURE) approach proposed in \cite{Ting2009}
experienced instability especially at high signal-to-noise ratios
(SNR). In the adopted Bayesian estimation framework, several
strategies are available to efficiently estimate these
hyperparameters in an unsupervised manner. Lavielle \emph{et al.}
proposed to couple Markov chain Monte Carlo (MCMC) methods to a
(stochastic) expectation-maximization (EM) algorithm
\cite{Lavielle2001,Kuhn2004}. A popular alternative to this hybrid
strategy consists of introducing a second level of hierarchy in the
Bayesian model by assigning non-informative prior distributions to
the unknown hyperparameters \cite[p. 383]{Robert2004}. The joint
posterior distribution of the unknown model parameters and
hyperparameter is then approximated from samples generated by MCMC
methods. This fully Bayesian estimation technique, followed in this
paper, has been recently applied to signal segmentation
\cite{Punskaya2002} and hyperspectral imaging
\cite{Dobigeon2008,Dobigeon2009sp}.

Besides, standard MCMC methods have shown some limitations for
deconvolving BG processes. More precisely, as noticed in
\cite{Bourguignon2005ssp} and \cite{Ge2008}, a standard Gibbs
sampler can be stucked in a particular configuration of the BG
process to be recovered, leading to poor mixing properties. Ge and
Idier recently demonstrated that this BG deconvolution can be easily
improved by marginalizing over the amplitudes of the non-zero
components \cite{Ge2008}. The resulting MCMC scheme is a partially
collapsed Gibbs sampler deeply studied by van Dyk \emph{et al.} in
\cite{vanDyk2008} and \cite{Park2009}. Following this approach, we
propose in this paper to take advantage of this MCMC strategy to
estimate the sparse sources efficiently.

To avoid redundant atoms in the dictionary and, more generally,
ensure a full rank mixing matrix, we address the blind source
separation problem under orthogonality constraint on the mixing
matrix. The main motivation for imposing this orthogonality
constraint on the mixing matrix is to capture more diversity among
the recovered atoms belonging to the under-complete dictionary to be
estimated. Only a few works in the signal processing literature have
considered this additional property. Preliminary results on this
issue have been reported in that has addressed the problem of sparse
source separation from orthogonal mixtures. However, the strategy
was based on a strong hypothesis for the source and mixing matrix,
i.e., the prior knowledge of the sparsity level shared by all the
sources. Hoff recently proposed in \cite{Hoff2007} a Bayesian
formulation of the dimension reduction operators. More precisely, to
estimate the rank of an unobserved matrix $\bfX$ involved in a noisy
model, he has derived a Bayesian description of the singular value
decomposition (SVD). The idea is to decompose the unobserved
noise-free data $\bfX$ as $\bfX=\bfU\bfD\bfV\transp$ where $\bfU$
and $\bfV$ are matrices with orthogonal columns. The Bayesian
inference on $\bfU$ and $\bfV$ has been finally conducted after
assigning uniform prior distributions for $\bfU$ and $\bfV$ on their
definition space, called the Stiefel manifold. This choice, coupled
with the Gaussian properties of the noise, leads to von Mises-Fisher
conditional posterior distributions for the columns of the matrices
$\bfU$ and $\bfV$. In the Bayesian orthogonal component analysis
(BOCA) studied in this paper, a similar strategy is adopted by
assigning a uniform distribution on the Stiefel manifold manifold to
the mixing matrix. The resulting MCMC algorithm generates mixing
matrix samples distributed according to the posterior distribution
following the efficient scheme developed in \cite{Hoff2007}.

This paper is organized as follows. The BOCA is formulated as a
blind source separation problem under constraints in
Section~\ref{sec:problem}. Section~\ref{sec:bayesian_model} derives
the statistical quantities required to define the Bayesian model.
The BOCA MCMC algorithm is described step-by-step in
Section~\ref{sec:Gibbs}. This algorithm allows one to generate
samples distributed according to the joint posterior distribution of
the source and mixing matrices. Simulation results conducted on
synthetic data, as well as a performance comparison with the K-SVD
algorithm, are reported in Section~\ref{sec:simulations}. Section
\ref{sec:image_proc} illustrates the interest of the proposed
algorithm by solving a sparse coding problem with an application to
natural image processing. Conclusions and potential future works are
considered in Section~\ref{sec:conclusion}.

\section{Problem formulation}
\label{sec:problem} Let
$\Vobs{\noobs}=\left[\obs{1}{\noobs},\ldots,\obs{\dimobs}{\noobs}\right]\transp$
denote measurement vectors of $\R^{\dimobs}$ observed at time
instants $t=1,\ldots,\nbobs$ by $\dimobs$ sensors. These
observations are assumed to be related to $\nbsou < \dimobs$
unobserved sources denoted $
\Vsou{\noobs}=\left[\sou{1}{\noobs},\ldots,\sou{\nbsou}{\noobs}\right]\transp$
via the
 matrix $\MATaxe$
in the following noisy linear model
\begin{equation}
    \label{eq:model}
  \Vobs{\noobs} = \MATaxe\Vsou{\noobs} + \Vnoise{\noobs}
\end{equation}
where $\Vnoise{\noobs}$ stands for an additive measurement noise.
Standard matrix notations yield
\begin{equation}
\label{eq:model_matrix}
  \MATobs = \MATaxe\MATsou + \MATnoise
\end{equation}
with $\MATobs=\left[\Vobs{1},\ldots,\Vobs{\nbobs}\right]$,
$\MATsou=\left[\Vsou{1},\ldots,\Vsou{\nbobs}\right]$ and
$\MATnoise=\left[\Vnoise{1},\ldots,\Vnoise{\nbobs}\right]$. The
$\dimobs\times 1$ noise vectors $\Vnoise{\noobs}$
($\noobs=1,\ldots,\nbobs$) are assumed to be independent and
distributed according to a centered multivariate Gaussian
distribution
$\calN\left(\Vzero{\dimobs},\noisevar\Id{\dimobs}\right)$.

In this work, the $\dimobs \times \nbsou$ matrix $\MATaxe$ is
assumed to be an unknown orthogonal matrix
\begin{equation}
  \Vaxe{i}\transp\Vaxe{j} = \left\{
                                        \begin{array}{ll}
                                          1, & \hbox{if }i= j \\
                                          0, & \hbox{if }i\neq j
                                        \end{array}
                                      \right.
\end{equation}
where the sources to be recovered can be sparsely represented.
Consequently, since only a few sources are assumed to be active at
time index $t$, the unobserved vector of $\nbsou$ sources
$\Vsou{\noobs}$ is sparse and contains only a few components that
are non-zero.

This paper proposes a Bayesian model as well as an MCMC sampling
strategy to estimate the unknown sources $\MATsou$, the orthogonal
matrix $\MATaxe$ and the noise variance $\noisevar$.

\section{Bayesian model}

\label{sec:bayesian_model} The unknown parameter vector associated
with the mixing model defined in \eqref{eq:model} is
$\paramvect=\left\{\MATsou,\MATaxe,\noisevar\right\}$. This section
gives the likelihood function of the observations and introduces
prior distributions for the unknown model parameters (assumed to be
\emph{a priori} independent).

\subsection{Likelihood function} The
Gaussian property of the additive noise yields for each observed
vector $\Vobs{\noobs}$
\begin{equation}
  f\left(\Vobs{\noobs}|\MATaxe,\Vsou{\noobs},\noisevar\right) =
\left(\frac{1}{2\pi\noisevar}\right)^{\frac{\dimobs}{2}}
\exp\left[-\frac{1}{2\noisevar}\norm{\Vobs{\noobs}-\MATaxe\Vsou{\noobs}}^2\right]
\end{equation}
where $\noobs=1,\ldots,\nbobs$ and $\norm{\cdot}$ stands for the
standard $\ell_2$-norm. By assuming the noise vectors
$\Vnoise{\noobs}$ to be \emph{a priori } independent, the full
likelihood function is
\begin{equation}
    \label{eq:likelihood}
  f\left(\MATobs|\MATaxe,\MATsou,\noisevar\right) =
\left(\frac{1}{2\pi\noisevar}\right)^{\frac{\nbobs\dimobs}{2}}\exp\left[-\frac{1}{2\noisevar}\sum_{\noobs=1}^{\nbobs}\norm{\Vobs{\noobs}-\MATaxe\Vsou{\noobs}}^2\right].
\end{equation}

\subsection{Noise variance prior}
As in numerous works, including
\cite{Punskaya2002,Dobigeon2007b,Dobigeon2008}, a conjugate
inverse-Gamma distribution is chosen as prior distribution for the
noise variance $\noisevar$ 
\begin{equation}
  \noisevar| \gamma \sim
\mathcal{IG}\left(\frac{\nu}{2},\frac{\gamma}{2}\right)
\end{equation}
where $\nu=2$ and $\gamma$ is an unknown hyperparameter that will be
estimated from the data. The main motivation for choosing conjugate
prior distribution for $\noisevar$ is to simplify the computation of
the posterior distribution of interest.

\subsection{Prior for the mixing matrix}
The mixing matrix $\MATaxe$ to be estimated is an
$\dimobs\times\nbsou$ matrix with orthogonal columns whose rank is
$\nbsou$. The set of such matrices, denoted
$\Steifel{\nbsou,\dimobs}$, is called the Stiefel
manifold\footnote{Note that for the special case $\dimobs=\nbsou$,
the Stiefel manifold $\Steifel{\nbobs,\nbobs}$ is the orthogonal
group $\calO\left(\dimobs\right)$ of orthogonal $\nbobs\times
\nbobs$ matrices.} (see \cite[p. 8]{Chikuse2003} for a general
introduction of this space). To reflect the absence of any
additional prior knowledge regarding the mixing matrix, a uniform
distribution on this set is chosen as prior distribution for
$\MATaxe$ \cite[p. 279]{Gupta2000}
\begin{equation}
\label{eq:prior_axes}
  f\left(\MATaxe\right) =
\frac{1}{\vol{\Steifel{\nbsou,\dimobs}}}
\Indicfun{\Steifel{\nbsou,\dimobs}}{\MATaxe}
\end{equation}
where $\Indicfun{\cdot}{\cdot}$ stands for the indicator function
\begin{equation}
  \Indicfun{\Steifel{\nbsou,\dimobs}}{\MATaxe} = \left\{
     \begin{array}{ll}
       1 & \hbox{if $\MATaxe\in \Steifel{\nbsou,\dimobs}$,} \\
       0 & \hbox{otherwise.}
     \end{array}
   \right.
\end{equation}
In \eqref{eq:prior_axes}, $\vol{\Steifel{\nbsou,\dimobs}}$ is the
volume of the Stiefel manifold $\Steifel{\nbsou,\dimobs}$ given by
\cite[p. 70]{Muirhead2005}
\begin{equation}
\vol{\Steifel{\nbsou,\dimobs}} = \frac{2^\dimobs
\pi^{\frac{\nbsou\dimobs}{2}}}{\Gamma_\dimobs\left(\frac{\nbsou}{2}\right)}
\end{equation}
where $\Gamma_\dimobs\left(\cdot\right)$ is the $\dimobs$-variate
Gamma function
\begin{equation}
\Gamma_\dimobs\left(u\right) =
\pi^{\frac{\dimobs\left(\dimobs-1\right)}{4}}\prod_{m=1}^{\dimobs}\Gamma\left(u+\frac{1-m}{2}\right)
\end{equation}
and $\Gamma\left(\cdot\right)$ is the Gamma function
\begin{equation}
\Gamma\left(u\right) = \int_0^{+\infty}t^{u-1}e^{-t}dt
\end{equation}
with $u>0$.

Generating samples according to \eqref{eq:prior_axes} can be easily
achieved by first sampling an $\dimobs \times \nbsou$ matrix $\bfV$
of independent standard normal random variables and then by setting
$\MATaxe = \bfV \left(\bfV\transp\bfV\right)^{-\frac{1}{2}}$
\cite{Chikuse2003}. However, as highlighted by Hoff in
\cite{Hoff2007}, sampling $\MATaxe$ via its conditional
distributions is frequently required, especially within an MCMC
estimation framework. Therefore, we recall below the procedure
proposed in \cite{Hoff2007} to sample orthogonal matrices $\MATaxe$
according to the uniform distribution \eqref{eq:prior_axes} using
the conditional distributions of its columns.

Firstly, let $\MATaxe_{\calA} = \left[\Vaxe{i}\right]_{i\in\calA}$
denote the matrix formed by the columns of $\MATaxe$ indexed by the
label vector $\calA \subset \left\{1,\ldots,\nbsou\right\}$, where
$\Vaxe{i}$ stands for the $i$th column of $\MATaxe$. Let
$\bsN_\calA$ denote an orthogonal basis associated with the null
space of the orthogonal matrix $\MATaxe_{\calA}$. Then, as
demonstrated in \cite{Hoff2007}, an orthogonal $\dimobs\times\nbsou$
matrix $\MATaxe$ can be uniformly drawn on the Stiefel manifold
$\Steifel{\nbsou,\dimobs}$ via the following steps
\begin{enumerate}
  \item Sample $\bsv_{1}$ uniformly on the unit
  $\dimobs$-sphere  and set $\Vaxe{1}=\bsv_1$,
  \item Sample $\bsv_{2}$ uniformly on the unit $(\dimobs-1)$-sphere
  and set $\Vaxe{2}=\bsN_1 \bsv_2$,
  \item Sample $\bsv_{3}$ uniformly on the unit $(\dimobs-2)$-sphere
  and set $\Vaxe{3}=\bsN_{\left\{1,2\right\}} \bsv_3$,
  \item[] $\vdots$
  \item[\nbsou)] Sample $\bsv_{\nbsou}$ uniformly on the unit $(\dimobs-\nbsou+1)$-sphere
  and set $\Vaxe{\nbsou}=\bsN_{\left\{1,\ldots,\nbsou-1\right\}}
  \bsv_\nbsou$.
\end{enumerate}
Uniform sampling on a sphere required in the scheme detailed above
can be easily achieved following the normal-deviate method described
in \cite{VonHohenbalken1980}. Finally, we have to mention that a
similar strategy will be used in Section~\ref{sec:Gibbs} to sample
mixing matrices $\MATaxe$ according to their conditional posterior
distributions.

\subsection{Source prior}
\label{subsec:prior_lambda} Since the source vectors $\Vsou{\noobs}$
are sparse, most of the elements $\sou{\nosou}{\noobs}$
($\nosou=1,\ldots,\nbsou$, $\noobs=1,\ldots,\nbobs$) of the matrix
$\MATsou$ are expected to be equal to zero. Therefore, choosing a
``sparse" prior for $\sou{\nosou}{\noobs}$ is recommended. Coupling
a standard probability density function (pdf) with an atom at zero
is a classical strategy to ensure sparsity. This strategy has been
widely used for located event detection \cite{Kormylo1982} such as
spike train deconvolution \cite{Champagnat1996, Cheng1996},
astrophysical frequency detection \cite{Bourguignon2005ssp}, sparse
approximations of times-series \cite{Blumensath2007} and
reconstruction of molecular images \cite{Dobigeon2009ip}. We propose
here to take advantage of this approach by choosing a BG
distribution as prior for $\sou{\nosou}{\noobs}$. The distribution
of this BG prior is defined as the following mixture
\begin{equation}
\label{eq:prior_sources}
  f\left(\sou{\nosou}{\noobs}|\souproba{\nosou},\souvar{\nosou}\right)
        = \left(1-\souproba{\nosou}\right)\dirac{\sou{\nosou}{\noobs}} +
            \souproba{\nosou} g_{\souvar{\nosou}}\left(\sou{\nosou}{\noobs}\right)
\end{equation}
where $\dirac{\cdot}$ is the Dirac delta function and
$g_{\souvar{\nosou}}\left(\sou{\nosou}{\noobs}\right)$ is the pdf of
the centered Gaussian distribution with variance $\souvar{\nosou}$.
In \eqref{eq:prior_sources}, the unknown hyperparameter
$\souproba{\nosou}$ is the prior probability of having an active
source. Consequently, this hyperparameter tunes the degree of
sparseness of the source vector $\Vsou{\noobs}$. Note that this
probability $\souproba{\nosou}$ of having an active source, as well
as the non-zero component variance $\souvar{\nosou}$, have been
assumed to be different from one source to another to provide a
flexible model. This strategy have been previously adopted in
\cite{Blumensath2007,Fevotte2007,Fevotte2008}. Another strategy
would be to assume that the sources have a non-homogeneous sparsity
level over times as explained in \cite{Dobigeon2009TR}. By assuming
that the source amplitudes $\sou{\nosou}{\noobs}$ are \emph{a
priori} independent, and introducing the index subsets
$\calI_{\nosou}(\epsilon) =
\left\{\noobs;\sou{\nosou}{\noobs}=\epsilon\right\}$ ($\epsilon \in
\left\{0,1\right\}$), the full prior distribution for the source
matrix $\MATsou$ is
\begin{equation}
\begin{split}
f\left(\MATsou|\Vsouproba,\Vsouvar\right) &=
\prod_{\nosou=1}^{\nbsou}\left[\left(1-\souproba{\nosou}\right)^{m_\nosou(0)}
    \prod_{\noobs\in \calI_\nosou(0)}
    \delta\left({\sou{\nosou}{\noobs}}\right)\right]\\
&\times\prod_{\nosou=1}^{\nbsou}\left[\souproba{\nosou}^{m_\nosou(1)}
 \prod_{\noobs
\in\calI_\nosou(1)}\left(\frac{1}{2\pi\souvar{\nosou}}\right)^{\frac{1}{2}}
\exp\left(-\frac{\sou{\nosou}{\noobs}^2}{\souvar{\nosou}}\right)\right],
\end{split}
\end{equation}
where
$\Vsouproba=\left[\souproba{1},\ldots,\souproba{\nosou}\right]\transp$,
$\Vsouvar=\left[\souvar{1},\ldots,\souvar{\nosou}\right]\transp$ and
$m_\nosou(\epsilon) =
\mathrm{card}\left\{\calI_{\nosou}(\epsilon)\right\}$. Note that
$m_\nosou(1) = \norm{\bfs_{\nosou}\transp}_0$, where
$\bfs_{\nosou}=\left[\sou{\nosou}{1},\ldots,\sou{\nosou}{\nbobs}\right]$
and $\norm{\cdot}_0$ denotes the $\ell_0$-norm, is the number of
active components in the source $\nosou$, whereas $m_\nosou(0) =
\nbobs-m_\nosou(1)$ is the number of components that are equal to
$0$ in the source $\nosou$.

\subsection{Hyperparameter priors}
A non-informative Jeffreys' prior is elected as prior distribution
for the hyperparameter $\gamma$
\begin{equation}
  f\left(\gamma\right) \propto
  \frac{1}{\gamma}\Indicfun{\R+}{\gamma}.
\end{equation}
An inverse-gamma distribution with fixed hyper-hyperparameters
$\alpha_0$ and $\alpha_1$ (in order to obtain vague prior with large
variance) is chosen as prior for the variance $\souvar{\nosou}$ of
the non-zero components in each source
\begin{equation}
  \souvar{\nosou} \sim \calI\calG\left(\alpha_0,\alpha_1\right).
\end{equation}
A uniform distribution is chosen as prior distribution for the
$\souproba{\nosou}$ in each source
\begin{equation}
\label{eq:prior_lambda}
  \souproba{\nosou} \sim \calU\left(\left[0,1\right]\right).
\end{equation}

Assuming that the individual hyperparameters are independent the
full prior distribution for the hyperparameter vector
$\hypervect=\left\{\gamma,\Vsouproba,\Vsouvar\right\}$ can be
expressed as
\begin{equation}
\label{eq:prior_hyper}
  f\left(\hypervect\right) \propto
\frac{1}{\gamma}\Indicfun{\R+}{\gamma}
\prod_{\nosou=1}^{\nbsou}\left[\left(\frac{1}{\souvar{\nosou}}\right)^{\alpha_0+1}\exp\left(-\frac{\alpha_1}{\souvar{\nosou}}\right)
\Indicfun{\left[0,1\right]}{\souproba{\nosou}}\right].
\end{equation}

\subsection{Posterior distribution}

\begin{figure}[h!]
\centering
  \includegraphics[width=\figwidth]{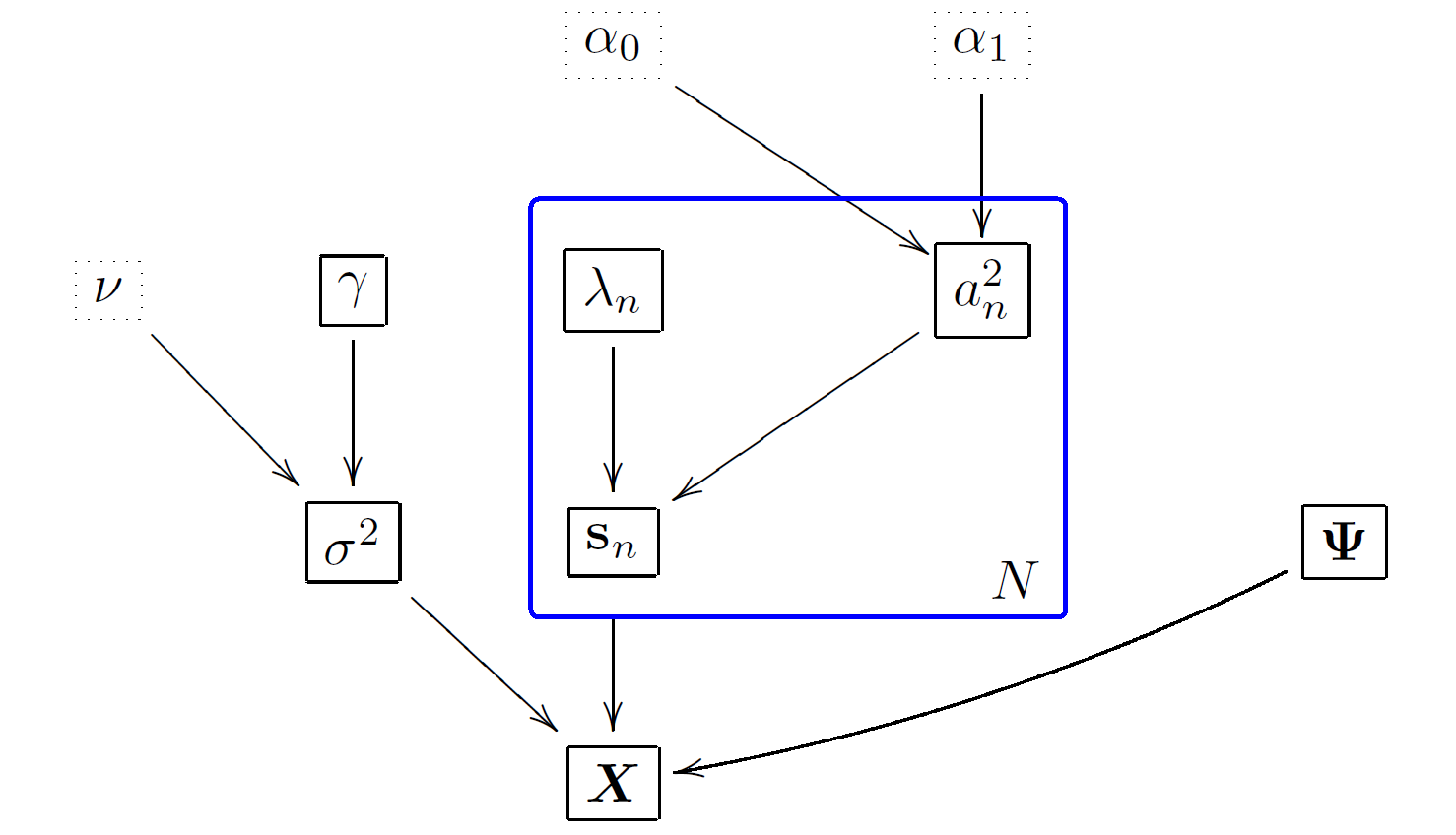}
\caption{DAG for the parameter priors and hyperpriors (the fixed
hyperparameters appear in dashed boxes).}\label{fig:DAG}
\end{figure}

The posterior distribution of $\left\{\paramvect,\hypervect\right\}$
can be computed from the following hierarchical structure
\begin{equation}
\label{eq:fullposterior} f\left(\paramvect,\hypervect|\MATobs\right)
\propto
f\left(\MATobs|\paramvect\right)f\left(\paramvect|\hypervect\right)f\left(\hypervect\right)
\end{equation}
where $\propto$ means proportional to,
\begin{equation}
f\left(\paramvect|\hypervect\right) =
f\left(\MATsou|\Vsouproba,\Vsouvar\right)f\left(\MATaxe\right)f\left(\noisevar|
\gamma\right)
\end{equation}
and where $f\left(\MATobs|\paramvect\right)$ and
$f\left(\hypervect\right)$ have been defined in
\eqref{eq:likelihood} and \eqref{eq:prior_hyper}. This hierarchical structure is represented as a graphical model on the directed acyclic
graph (DAG) of Fig. \ref{fig:DAG}. In the joint
distribution \eqref{eq:fullposterior}, the nuisance parameter
$\gamma$ can be easily integrated out, leading to
\begin{equation}
\begin{split}
 \label{eq:posterior}
  f&\left(\MATsou,\MATaxe,\noisevar,\Vsouproba,\Vsouvar|\MATobs\right)\propto
  \Indicfun{\Steifel{\nbsou,\dimobs}}{\MATaxe}\left(\frac{1}{\noisevar}\right)^{\frac{\nbobs\dimobs + \nu}{2}}
        \exp\left[-\frac{1}{2\noisevar}\sum_{\noobs=1}^{\nbobs}\norm{\Vobs{\noobs}-\MATaxe\Vsou{\noobs}}^2\right]\\
&\times\prod_{\nosou=1}^{\nbsou}\left[\left(1-\souproba{\nosou}\right)^{m_\nosou(0)}
    \prod_{\noobs\in \calI_\nosou(0)}
    \delta\left({\sou{\nosou}{\noobs}}\right)\right]\prod_{\nosou=1}^{\nbsou}\left[\souproba{\nosou}^{m_\nosou(1)}
 \prod_{\noobs
\in\calI_\nosou(1)}\left(\frac{1}{2\pi\souvar{\nosou}}\right)^{\frac{1}{2}}
\exp\left(-\frac{\sou{\nosou}{\noobs}^2}{\souvar{\nosou}}\right)\right]\\
&\times
\prod_{\nosou=1}^{\nbsou}\left[\left(\frac{1}{\souvar{\nosou}}\right)^{\alpha_0+1}\exp\left(-\frac{\alpha_1}{\souvar{\nosou}}\right)
\Indicfun{\left[0,1\right]}{\souproba{\nosou}}\right].
\end{split}
\end{equation}
Inferring the source matrix $\MATsou$ and the orthogonal matrix
$\MATaxe$ from \eqref{eq:posterior} is not straightforward, mainly
due to the combinatory problem induced by the quantities $n_1(t)$
and $n_0(t)$. In particular, closed-form expressions of the Bayesian
estimators of $\MATsou$ and $\MATaxe$ are difficult to obtain. We
propose to use MCMC methods to generate samples that are
asymptotically distributed according to the target distribution
\eqref{eq:posterior}. These generated samples are then used to
approximate the Bayesian estimators of $\MATsou$ and $\MATaxe$.

\section{Partially collapsed Gibbs sampler for orthogonal component analysis of sparse sources}
\label{sec:Gibbs} We describe in this section an MCMC method that
allows one to generate a sample collection
$$\mathcal{Y} =\left\{\left(\sample{\MATsou}{h},\sample{\MATaxe}{h},\samplenoisevar{h},\sample{\Vsouproba}{h},\sampleVsouvar{h}\right)\right\}_{h=1,\ldots,N_{\mathrm{MC}}}$$
asymptotically distributed according to the posterior distribution
\eqref{eq:posterior}. The interested reader is invited to consult
\cite{Robert2004} for more details about MCMC methods.

The easiest way to sample according to this posterior would consist
of using a standard Gibbs sampler whose main steps are
\begin{enumerate}
  \item sample $\MATsou$ from   $f\left(\MATsou  |\MATaxe,\noisevar,\Vsouproba,\Vsouvar,\MATobs\right)$,
  \item sample $\MATaxe$ from   $f\left(\MATaxe  |\MATsou,\noisevar,\Vsouproba,\Vsouvar,\MATobs\right) = f\left(\MATaxe|\MATsou,\noisevar,\MATobs\right)$
  \item sample $\noisevar$ from $f\left(\noisevar|\MATsou,\MATaxe,\Vsouproba,\Vsouvar,\MATobs\right) = f\left(\noisevar|\MATsou,\MATaxe,\MATobs\right)$,
  \item sample $\Vsouproba$ from   $f\left(\Vsouproba  |\MATsou,\MATaxe,\noisevar,\Vsouvar,\MATobs\right) = f\left(\Vsouproba|\MATsou\right)$,
  \item sample $\Vsouvar$ from   $f\left(\Vsouvar  |\MATsou,\MATaxe,\noisevar,\Vsouproba,\MATobs\right) = f\left(\Vsouvar|\MATsou\right)$.
\end{enumerate}
However, as highlighted in previous works, sampling BG processes
following the crude Gibbs sampler detailed above [step 1)] often
leads to poor mixing properties and weak estimation performance
\cite{Bourguignon2005ssp}. As an alternative a new MCMC algorithm
for BG deconvolution was recently studied \cite{Ge2008}. The
approach relies on explicitly introducing binary variables
$\MATindic$ that indicate the presence of non-zero BG components.
Then, these indicators are sampled after marginalizing over the BG
variable amplitudes. This strategy casts the resulting MCMC
algorithm as a partially collapsed Gibbs (PCG) sampler. Van Dyk and
Park have described in \cite{Park2009} and \cite{vanDyk2008} how PCG
samplers can be efficient tools to overcome drawbacks inherent to
standard Gibbs sampler, e.g., slow convergence. As detailed in the
works cited above, PCG samplers consists of replacing some of the
conditional distributions with marginalized conditional
distribution. The resulting PCG sampling scheme can be summarized by
the following steps
\begin{enumerate}
  \item sample $\MATindic$ from $f\left(\MATindic|\MATaxe,\noisevar,\Vsouproba,\Vsouvar,\MATobs\right)$,
  \item sample $\MATsou$ from $f\left(\MATsou|\MATindic,\MATaxe,\noisevar,\Vsouproba,\Vsouvar,\MATobs\right)$,
  \item sample $\MATaxe$ from  $f\left(\MATaxe|\MATsou,\noisevar,\MATobs\right)$
  \item sample $\noisevar$ from $f\left(\noisevar|\MATsou,\MATaxe,\MATobs\right)$,
  \item sample $\Vsouproba$ from $f\left(\Vsouproba|\MATsou\right)$,
  \item sample $\Vsouvar$ from $f\left(\Vsouvar|\MATsou\right)$.
\end{enumerate}
Note that the source amplitudes have been marginalized to provide
the discrete distribution appearing in step 1). The main steps of
the PCG sampler are detailed in
subsections~\ref{subsec:sample_sources}
to~\ref{subsec:sample_souvar} (see also the step-by-step
Algo.\ref{algo:Gibbs}).

\begin{algorithm}[h!]
    \caption{Gibbs sampling for orthogonal component analysis of sparse sources}
    \label{algo:Gibbs}
    \begin{algorithmic}[1]
        \STATE \emph{\scriptsize{\% sampling the sources}}
         \FOR{$\noobs=1~\text{to}~\nbobs$}
            \STATE \emph{\scriptsize{\% sampling the indicators recursively following Algo. \ref{algo:recursive_indic}}}
            \FOR{$\nosou=1~\text{to}~\nbsou$}
            \STATE sample the indicator $\indic{\nosou}{\noobs}$ following the probability \eqref{eq:posterior_proba_indic},
            \ENDFOR
            \STATE sample the source vector $\Vsou{\noobs}$ from the pdf's in \eqref{eq:posterior_sources1} and \eqref{eq:posterior_sources2},
         \ENDFOR
         \STATE \emph{\scriptsize{\% sampling the orthogonal mixing matrix}}
        \FOR{$\nosou=1~\text{to}~\nbsou$}
            \STATE compute the basis $\bsN_n$ of the null space of $\MATaxe_{-n}$,
            \STATE sample $\bsv_{\nosou}$ from the von Mises-Fisher distribution in
                    \eqref{eq:posterior_v},
            \STATE set $\Vaxe{\nosou} =  \bsN_{\nosou}\bsv_{\nosou}$,
        \ENDFOR
         \STATE \emph{\scriptsize{\% sampling the noise variance}}
         \STATE sample parameter $\noisevar$ from the pdf in \eqref{eq:posterior_noisevar},
        \STATE \emph{\scriptsize{\% sampling the probability of having active sources}}
        \FOR{$\nosou=1~\text{to}~\nbsou$}
        \STATE sample the hyperparameter $\souproba{\nosou}$ from the pdf in
            \eqref{eq:posterior_lambda},
        \STATE \emph{\scriptsize{\% sampling the active source variances}}
        \STATE sample the hyperparameter $\souvar{\nosou}$ from the pdf in
            \eqref{eq:posterior_magnparam},
        \ENDFOR
    \end{algorithmic}
\end{algorithm}

\subsection{Sampling the indicator and source matrices}
\label{subsec:sample_sources} The prior independence assumption of
the source vectors allows one to rewrite the joint posterior
distribution of the source matrix $\MATsou$ as
\begin{equation}
  f\left(\MATsou|\Vsouproba,\Vsouvar,\noisevar,\MATaxe,\MATobs\right) =
\prod_{\noobs=1}^{\nbobs}
f\left(\Vsou{\noobs}|\Vsouproba,\Vsouvar,\noisevar,\MATaxe,\Vobs{\noobs}\right).
\end{equation}
Consequently, sampling according to
$f\left(\MATsou|\Vsouproba,\Vsouvar,\noisevar,\MATaxe,\MATobs\right)$
can be achieved by successively sampling the source vectors
$\Vsou{\noobs}$ according to
$f\left(\Vsou{\noobs}|\Vsouproba,\Vsouvar,\noisevar,\MATaxe,\Vobs{\noobs}\right)$
for $\noobs=1,\dots,\nbobs$.\\

It is important to note here that standard derivations similar to
those in \cite{Cheng1996} allow one to state that the conditional
posterior distribution of the $\noobs$th component
$\sou{\nosou}{\noobs}$ of the $\nosou$th source is a BG
distribution. However, as pointed out in \cite{Bourguignon2005ssp},
sampling according to this distribution needs to explore the state
space efficiently, which can be difficult mainly due to the
difficulty of the Gibbs sampler to escape from local maxima.
Recently, Ge \emph{et al} have introduced a performing MCMC
algorithm to overcome this issue by explicitly introducing an
auxiliary binary variable $\indic{\nosou}{\noobs}$ that indicates
the active sources \cite{Ge2008}
\begin{equation}
  \indic{\nosou}{\noobs} = \left\{
    \begin{array}{ll}
        1, & \hbox{if $\sou{\nosou}{\noobs}\neq0$,} \\
        0, & \hbox{otherwise.}
    \end{array}
    \right.
\end{equation}
Conditionally upon this indicator variable $\indic{\nosou}{\noobs}
$, the prior of the source component $\sou{\nosou}{\noobs} $ in
\eqref{eq:prior_sources} can be easily rewritten
\begin{align*}
  f\left(\sou{\nosou}{\noobs}|\indic{\nosou}{\noobs}=0\right) &=
\delta\left(\sou{\nosou}{\noobs}\right),\\
  f\left(\sou{\nosou}{\noobs}|\indic{\nosou}{\noobs}=1,a\right) &=
g_{\souvar{\nosou}}\left(\sou{\nosou}{\noobs}\right).
\end{align*}
The probability of having an active component in source $\nosou$ is
governed by an unknown hyperparameter $\souproba{\nosou}$ such that
\begin{align*}
  \proba{\indic{\nosou}{\noobs}=1} &= \souproba{\nosou},\\
  \proba{\indic{\nosou}{\noobs}=0} &= 1-\souproba{\nosou}.
\end{align*}

In \cite{Ge2008}, Ge \emph{et al.} have proposed to sample the
source vectors $\Vsou{\noobs}$ ($\noobs=1,\ldots,\nbobs$) and the
indicators
$\Vindic{\noobs}=\left[\indic{1}{\noobs},\ldots,\indic{\nbsou}{\noobs}\right]\transp$
using the following $2$ steps
\begin{itemize}
  \item[1)] Sampling according to
$f\left(\Vindic{\noobs}|\Vsouproba,\Vsouvar,\noisevar,\MATaxe,\Vobs{\noobs}\right)$,
  \item[2)] Sampling according to
$f\left(\Vsou{\noobs}|\Vindic{\noobs},\Vsouproba,\Vsouvar,\noisevar,\MATaxe,\Vobs{\noobs}\right)$.
\end{itemize}
As mentioned above, these two steps make the resulting Gibbs sampler
a PCG sampler and are detailed below.

\subsubsection{Sampling according to
$f\left(\Vindic{\noobs}|\Vsouproba,\Vsouvar,\noisevar,\MATaxe,\Vobs{\noobs}\right)$}


Sampling  according to
$f\left(\Vindic{\noobs}|\Vsouproba,\Vsouvar,\noisevar,\MATaxe,\Vobs{\noobs}\right)$
can be performed by updating the $\nbsou$ components
$\indic{\nosou}{\noobs}$ ($\nosou=1,\ldots,\nbsou$) successively. As
noticed in \cite{Ge2008}, the posterior probability of having the
source component $\indic{\nosou}{\noobs}$ to be active given the
other components denoted $\VindicP{\noobs}{-\nosou}=
\left[\indic{1}{\noobs},\ldots,\indic{\nosou-1}{\noobs},\indic{\nosou+1}{\noobs},\ldots,\indic{\nbsou}{\noobs}\right]\transp$
is
\begin{equation}
\label{eq:posterior_proba_indic}
  \proba{\indic{\nosou}{\noobs}=1|\VindicP{\noobs}{-\nosou},\souproba{\nosou},\souvar{\nosou},\MATaxe,\noisevar,\Vobs{\noobs}}
=\left[1 + \exp\left(-\frac{u_0-u_1}{2}\right)\right]^{-1}
\end{equation}
with
\begin{align*}
  u_\epsilon &=  \Vobs{\noobs}\transp \bsB_\epsilon\inv\Vobs{\noobs}+
\log\left|\bsB_\epsilon\right|+ 2 \epsilon
\log\left(\frac{1}{\souproba{\nosou}}-1\right),\\
 \bsB_\epsilon &= \souvar{\nosou}\MATaxe\mathrm{diag}\left\{\VindicP{\noobs}{[\epsilon]}\right\}\MATaxe\transp +
\noisevar \Id{\dimobs},\\
  \VindicP{\noobs}{[\epsilon]} &=
\left[\indic{1}{\noobs},\ldots,\indic{\nosou-1}{\noobs},\epsilon,\indic{\nosou+1}{\noobs},\ldots,\indic{\nbsou}{\noobs}\right]\transp.
\end{align*}

The probability in \eqref{eq:posterior_proba_indic} can be
efficiently computed following the recursive scheme initially
introduced in \cite{Champagnat1996} (and used in \cite{Ge2008}), and
adapted here to take into account the orthogonality property of
$\MATaxe$. This numerical implementation relies on the Cholesky
decomposition of $\bsB_\epsilon$ and the matrix inversion lemma.
This avoids to calculate the compute-intensive inversion of
$\bsB_\epsilon$ and the determinant $\left|\bsB_\epsilon\right|$ at
each step of the Gibbs sampler. We describe in
Algo.~\ref{algo:recursive_indic} how the component
$\indic{\nosou}{\noobs}$ is updated.\\

\begin{algorithm}[h!]
    \caption{Recursive sampling of indicator vector $\Vindic{\noobs}$}
    \label{algo:recursive_indic}
        \begin{algorithmic}[1]
            \FOR{$\nosou=1~\text{to}~\nbsou$}
                \STATE set $\delta_\nosou = (-1)^{\indic{\nosou}{\noobs}}$,
                \STATE set $\bsG =
                    \left[\Vaxe{i}\right]_{\indic{i}{\noobs}=1}$,
                \STATE set $\mu_\nosou = \frac{\souvar{\nosou}}{\noisevar}$,
                \STATE set $\tau_\nosou = \delta_\nosou +
                    \mu_\nosou \norm{\Vaxe{\nosou}}^2-\frac{\mu_\nosou^2}{1+\mu_\nosou}\Vaxe{\nosou}\transp \bsG \bsG\transp
                    \Vaxe{\nosou}$,
                \STATE set $\eta_\nosou = \Vobs{\noobs}\transp\Vaxe{\nosou}-\frac{\mu_\nosou}{1+\mu_\nosou} \Vobs{\noobs}\transp \bsG \bsG\transp
                    \Vaxe{\nosou}$,
                \STATE set $\Delta u = \log(\delta_\nosou \tau_\nosou) - \frac{\mu_\nosou}{\noisevar \tau_\nosou} \eta_\nosou^2 + 2\delta_\nosou
                        \log\left(\frac{1}{\souproba{\nosou}}-1\right)$,
                \STATE sample $w\sim\calU_{[0,1]}$,
                \IF{$u>\left[1+\exp\left(-\frac{\Delta u}{2}\right)\right]^{-1}$}
                    \STATE set $\indic{\nosou}{\noobs} = \indic{\nosou}{\noobs} +
                        \delta_{\nosou}$,
                \ENDIF
            \ENDFOR
        \end{algorithmic}
\end{algorithm}

\subsubsection{Sampling according to
$f\left(\Vsou{\noobs}|\Vindic{\noobs},
\Vsouvar,\noisevar,\MATaxe,\Vobs{\noobs}\right)$}


Conditionally upon the indicator variable $\indic{\nosou}{\noobs}$,
the distribution of $\sou{\nosou}{\noobs}$ is defined by
\begin{equation}
\label{eq:posterior_sources1}
  f\left(\sou{\nosou}{\noobs}|\indic{\nosou}{\noobs}=0,\noisevar,\MATaxe,\Vobs{\noobs}\right)
= \delta\left(\sou{\nosou}{\noobs}\right)
\end{equation}
and
\begin{equation}
\label{eq:posterior_sources2}
 \left[\sou{n}{\noobs}\right]_{\indic{n}{\noobs}=1}|\Vindic{\noobs},\Vsouvar,\noisevar,\MATaxe,\Vobs{\noobs}
        \sim \calN\left(\boldsymbol{\Lambda}_1\bsG\Vobs{\noobs}, \boldsymbol{\Lambda}_2\right)
\end{equation}
where $\left[\sou{n}{\noobs}\right]_{\indic{n}{\noobs}=1}$ stands
for $L\times 1$ vector composed of the active components in the
source vector $\Vsou{\noobs}$, $L= \norm{\Vindic{\noobs}}_0$,
$\bsG=\left[\Vaxe{\nosou}\right]_{\indic{\nosou}{\noobs}=1}$ is the
$\dimobs\times L$ matrix composed of the columns of $\MATaxe$
corresponding to the active source components and
$\boldsymbol{\Lambda}_1 =
\mathrm{diag}\left\{\frac{\mu_\nosou}{\mu_\nosou+1}\right\}_{\indic{\nosou}{\noobs}=1}$
and $\boldsymbol{\Lambda}_2 =
\mathrm{diag}\left\{\frac{1}{\mu_\nosou+1}\right\}_{\indic{\nosou}{\noobs}=1}$
are $L\times L$ diagonal matrices with
$\mu_\nosou=\frac{\souvar{\nosou}}{\noisevar}$. Note that this block
sampling strategy, also adopted in \cite{Fevotte2007}, avoids to
sample the non-zero source components one-by-one.

\subsection{Sampling the mixing matrix}
 \label{subsec:sample_MATdir} The conditional distribution
$f\left(\MATaxe|\MATsou,\noisevar,\MATobs\right)$ being intractable,
this section describes how Gibbs moves can be used to generate
samples $\sample{\Vaxe{\nosou}}{m}$ according to the posterior
distribution of each column of $\MATaxe$ conditionally upon the
others. Let $\MATaxe_{-\nosou} =
\left[\Vaxe{i}\right]_{i\neq\nosou}$ (resp.
$\VsouP{-\nosou}{\noobs}$) denote the matrix $\MATaxe$ (resp. vector
$\Vsou{\noobs}$) whose $\nosou$th column (resp. component) has been
removed. By denoting
\begin{equation}
 \boldsymbol{\mu}_{\nosou}\left(\noobs\right)=\sou{\nosou}{\noobs}\left[ \Vobs{\noobs}-\MATaxe_{-\nosou}\VsouP{-\nosou}{\noobs}\right]
\end{equation}
straightforward computations yield
\begin{equation}
\label{eq:posterior_Vdir}
  f\left(\Vaxe{\nosou}|\MATaxe_{-\nosou},\MATsou,\noisevar,\MATobs\right)\propto
    \exp\left[\frac{1}{\noisevar}\sum_{\noobs=1}^{\nbobs}\boldsymbol{\mu}_\nosou(\noobs)\transp
\Vaxe{\nosou}\right] \Indicfun{\Steifel{\dimobs,\nbsou}}{\MATaxe}.
\end{equation}
As detailed in \cite{Hoff2007}, conditionally upon
$\MATaxe_{-\nosou}$, $\Vaxe{\nosou}$ can be written
$\Vaxe{\nosou}=\bsN_{\nosou}\bsv_{\nosou}$ where $\bsv_{\nosou}$ is
uniform on the sphere and $\bsN_{\nosou}$ is a basis for the null
space of $\MATaxe_{-\nosou}$. Therefore, from
\eqref{eq:posterior_Vdir}, the conditional distribution of
$\bsv_{\nosou}$ is
\begin{equation}
\label{eq:posterior_v}
  f\left(\bsv_{\nosou}|\MATaxe_{-\nosou},\MATsou,\noisevar,\MATobs\right)\propto
    \exp\left[\frac{1}{\noisevar}\sum_{\noobs}^{\nbobs}\boldsymbol{\mu}_{\nosou,\noobs}\transp
\bsN_{\nosou}\bsv_{\nosou}\right]
\end{equation}
which is a von Mises-Fisher distribution with parameter
$\frac{1}{\noisevar}\sum_{\noobs}^{\nbobs}\boldsymbol{\mu}_{\nosou,\noobs}\transp
\bsN_{\nosou}$. A standard method to sample according to this
distribution is given in \cite{Wood1994}. To summarize, the columns
of $\MATaxe$ can be iteratively sampled conditionally upon the
others by drawing samples $\bsv_{\nosou}$ from a von Mises-Fisher
distribution and setting $\Vaxe{\nosou} =
\bsN_{\nosou}\bsv_{\nosou}$.

\subsection{Sampling the noise variance}
\label{subsec:sample_noisevar} Looking carefully at
\eqref{eq:posterior}, the conditional distribution of the noise
variance is an inverse-gamma distribution such that
\begin{equation}
    \label{eq:posterior_noisevar}
  \noisevar\left|\MATsou,\MATaxe,\MATobs\right. \sim
\calI\calG\left(\frac{\nbobs\dimobs}{2},\frac{1}{2}\sum_{\noobs=1}^{\nbobs}\norm{\Vobs{\noobs}
- \MATaxe\Vsou{\noobs}}^2\right).
\end{equation}

\subsection{Sampling the probability of having an active source} The conditional
distribution of the hyperparameter $\souproba{\nosou}$ (i.e., the
probability of the source $\nosou$ to be active) can be computed
from \eqref{eq:posterior}
\begin{equation}
  f\left(\souproba{\nosou}|\bfs_\nosou\right) \propto
\left(1-\souproba{\nosou}\right)^{m_\nosou(0)}\souproba{\nosou}^{m_\nosou(1)}
\end{equation}
where $m_\nosou(\epsilon)$, $\epsilon\in\left\{0,1\right\}$, has
been defined in paragraph~\ref{subsec:prior_lambda}. Therefore,
sampling according to $f\left(\souproba{\nosou}|\bfs_\nosou\right)$
is achieved as follows
\begin{equation}
\label{eq:posterior_lambda}
 \souproba{\nosou}|\bfs_\nosou\sim
\mathcal{B} e\left(m_\nosou(0)+1,m_\nosou(1)+1\right).
\end{equation}
where $\calB e\left(a,b\right)$ is a Beta distribution with shape
parameters $a$ and $b$.

\subsection{Sampling the variance of the active sources}
\label{subsec:sample_souvar} Straightforward computations leads to
the following IG distribution as conditional posterior the variance
of non-zero BG components in the source $\nosou$
\begin{equation}
\label{eq:posterior_magnparam}
 \souvar{\nosou}|\bfs_\nosou \sim
\mathcal{IG}\left(\frac{1}{2}m_\nosou(1)+\alpha_0,\frac{1}{2}\norm{\bfs_\nosou\transp}^2+\alpha_1\right).
\end{equation}

\subsection{Inferring the sources and the mixing matrix}
\label{subsec:inferring} The main objective of the proposed Bayesian
algorithm is to estimate the source matrix $\MATsou$ and the mixing
matrix $\MATaxe$ from the data, independently from the nuisance
parameters $\noisevar$, $\souproba{\nosou}$ and $\souvar{\nosou}$.
The MCMC algorithm detailed in paragraphes
\ref{subsec:sample_sources} to \ref{subsec:sample_souvar} generates
samples asymptotically distributed according to the posterior
distribution \eqref{eq:posterior}. Consequently, the MMSE estimators
of $\MATsou$ and $\MATaxe$ can be approximated by empirical averages
over the $N_\mathrm{MC}-N_\mathrm{bi}$ drawn samples as follows
\begin{align} \label{eq:MMSE_estimate}
 \hat\MATsou_{\text{MMSE}} &= \mathrm{E}\left[\MATsou|\MATobs\right]  \approx
\frac{1}{N_\mathrm{MC}-N_\mathrm{bi}}\sum_{h=N_\mathrm{bi}+1}^{N_\mathrm{MC}}\sample{\MATsou}{h}\\
    \hat\MATaxe_{\text{MMSE}} &= \mathrm{E}\left[\MATaxe|\MATobs\right]  \approx
\frac{1}{N_\mathrm{MC}-N_\mathrm{bi}}\sum_{h=N_\mathrm{bi}+1}^{N_\mathrm{MC}}\sample{\MATaxe}{h}
\end{align}
where $N_\mathrm{bi}$ denotes the number of burn-in iterations of
the sampler and $N_\mathrm{MC}$ is the total number of Monte Carlo
iterations.

Another important property of the proposed MCMC algorithm is that
the generated pairs
$$\left\{\left(\sample{\MATsou}{1},\sample{\MATaxe}{1}\right),\ldots,\left(\sample{\MATsou}{N_\mathrm{MC}},\sample{\MATaxe}{N_\mathrm{MC}}\right)\right\}$$
form also a Markov chain whose stationary distribution is the joint
marginal distribution $f\left(\MATsou,\MATaxe|\MATobs\right)$.
Therefore, the joint MAP estimator of $\left(\MATsou,\MATaxe\right)$
can be computed by retaining among the collection
$$\mathcal{X} =
\left\{\left(\sample{\MATsou}{h},\sample{\MATaxe}{h}\right)
\right\}_{h=N_\mathrm{bi},\ldots,N_\mathrm{MC}}$$ the sample that
maximizes the marginalized distribution
$f\left(\MATsou,\MATaxe|\MATobs\right)$ \cite[p. 165]{Marin2007}
\begin{equation} \label{eq:MAP_estimate}
\begin{split} \left(\hat\MATsou_{\text{MAP}},\hat\MATaxe_{\text{MAP}}\right) &=
    \operatornamewithlimits{argmax}_{\left(\MATsou,\MATaxe\right)\in \mathbb{R}^{\nbsou\times\nbobs} \times \calS_{\dimobs,\nbsou}}
   f\left(\MATsou,\MATaxe|\MATobs\right)\\
    & \approx \operatornamewithlimits{argmax}_{\left(\MATsou,\MATaxe\right)\in \calX}
    f\left(\MATsou,\MATaxe|\MATobs\right).
\end{split}
\end{equation}
Note that this joint marginal distribution
$f\left(\MATsou,\MATaxe|\MATobs\right)$ can be easily computed from
the hierarchical structure \eqref{eq:fullposterior} that allows one
to integrate out the hyperparameter vector $\hypervect$ and the
noise variance $\noisevar$ in the full posterior
distribution~\eqref{eq:posterior}, yielding
\begin{equation}
\label{eq:posterior_marginal}
  f\left(\MATsou,\MATaxe|\MATobs\right) \propto
    \frac{\prod_{\nosou}^{\nbsou}B\left(1+ m_\nosou(1), 1 +
m_\nosou(0)\right)}{\left[\sum_{\noobs=1}^{\nbobs}\left\|\Vobs{\noobs}-\MATaxe\Vsou{\noobs}\right\|^2\right]^{\frac{\nbobs\dimobs}{2}}}
\prod_{\nosou}^{\nbsou}\frac{\Gamma\left(\alpha_0 +
\frac{m_\nosou(1)}{2}
\right)}{\left[\alpha_1+\frac{\norm{\bfs_\nosou\transp}^2}{2}\right]^{\alpha_0
+ \frac{ m_\nosou(1)}{2}}}
\end{equation}
where $B\left(a, b\right)=
\frac{\Gamma\left(a\right)\Gamma\left(b\right)}{\Gamma\left(a+b\right)}$.

\section{Simulation results}
\label{sec:simulations}
\subsection{Performance analysis} 
\begin{figure}[h!]
  \centering
  \includegraphics[width=\figwidth]{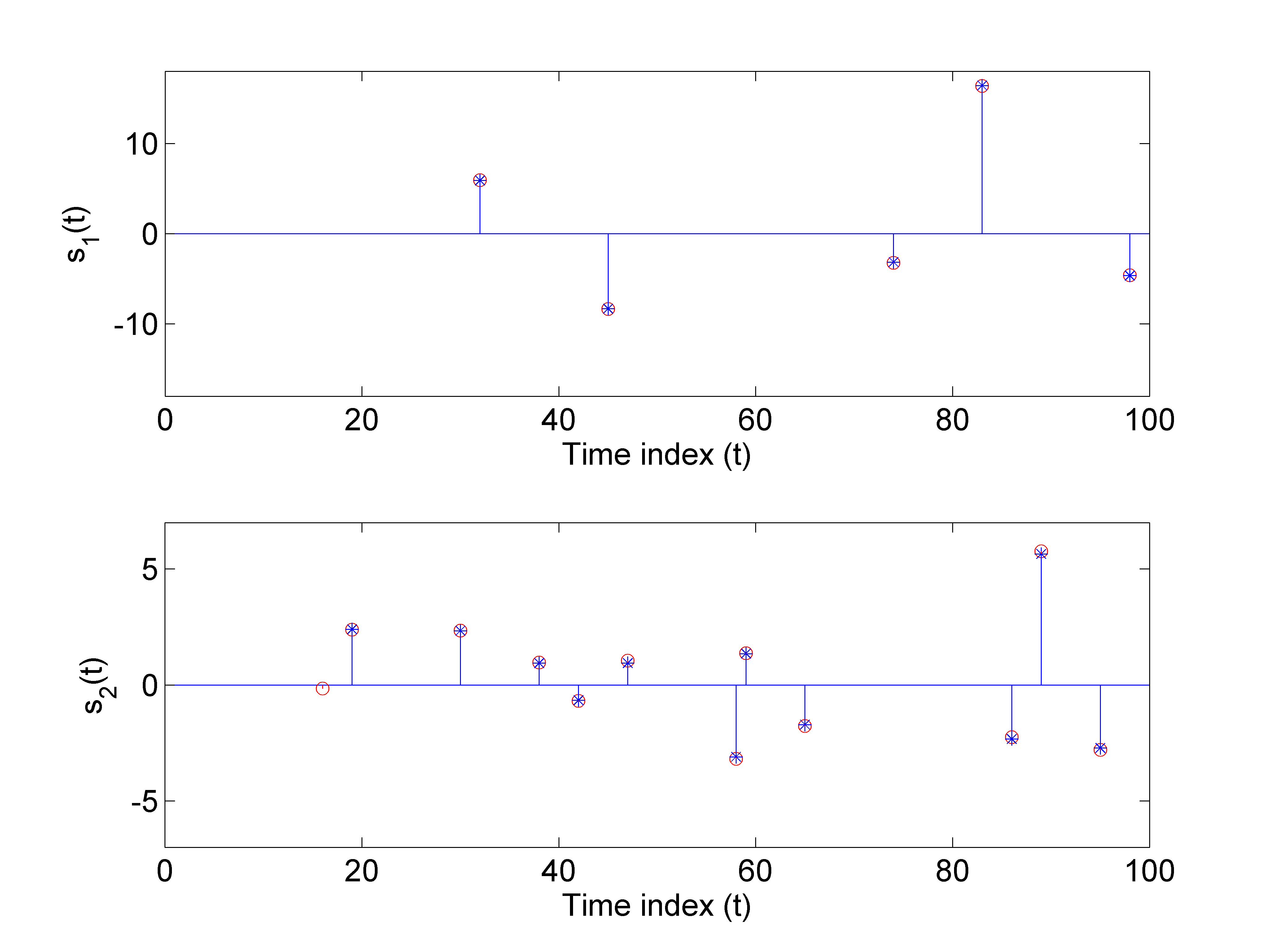}
  \caption{Actual sources (circles, red) and corresponding MAP estimates (stars, blue).}
  \label{fig:result_sources_SNR=15dB}
\end{figure}

This section first considers a toy example to provide comprehensive and extensive results.
We generate $\nbsou=2$ sources of length $\nbobs=100$ according to
the prior distribution in \eqref{eq:prior_sources} with the
probabilities of having active sources $\souproba{1}=0.05$,
$\souproba{2}=0.1$  and the active source variances $\souvar{1}=100$
and $\souvar{2}=10$. These $2$ sources, represented in
Fig.~\ref{fig:result_sources_SNR=15dB} (red), are mixed to obtain
$\nbobs=100$ observation vectors $\Vobs{\noobs}\in \mathbb{R}^{50}$
of dimension $\dimobs=50$.

The generated orthogonal mixing matrix $\MATaxe$ is composed of
$\nbsou=2$ basis vectors proportional to sinusoidal functions,
$\axe{m}{\nosou}\propto\cos\left(2\pi f_\nosou m + \pi\right)$
($m=1,\ldots,\dimobs$) with two different frequencies $f_1=0.02$ and
$f_2=0.04$. These two vectors are represented in
Fig.~\ref{fig:result_sources_SNR=15dB} (red).

\begin{figure}[h!]
  \centering
  \includegraphics[width=\figwidth]{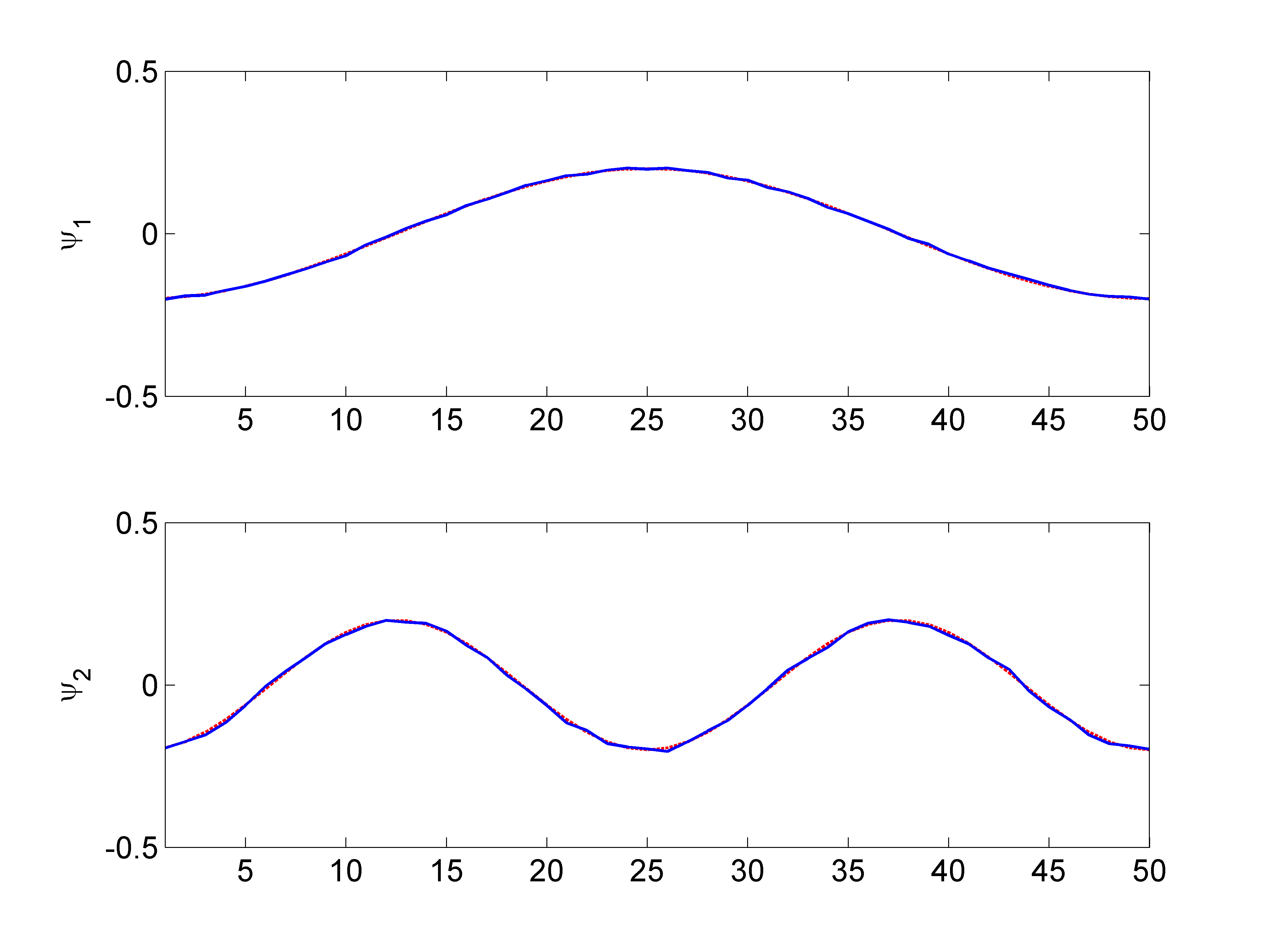}
  \caption{Actual basis vectors (red) and corresponding MAP estimates (blue).}
  \label{fig:result_psi_SNR=15dB}
\end{figure}

\begin{figure}[h!]
  \centering
  \includegraphics[width=\figwidth]{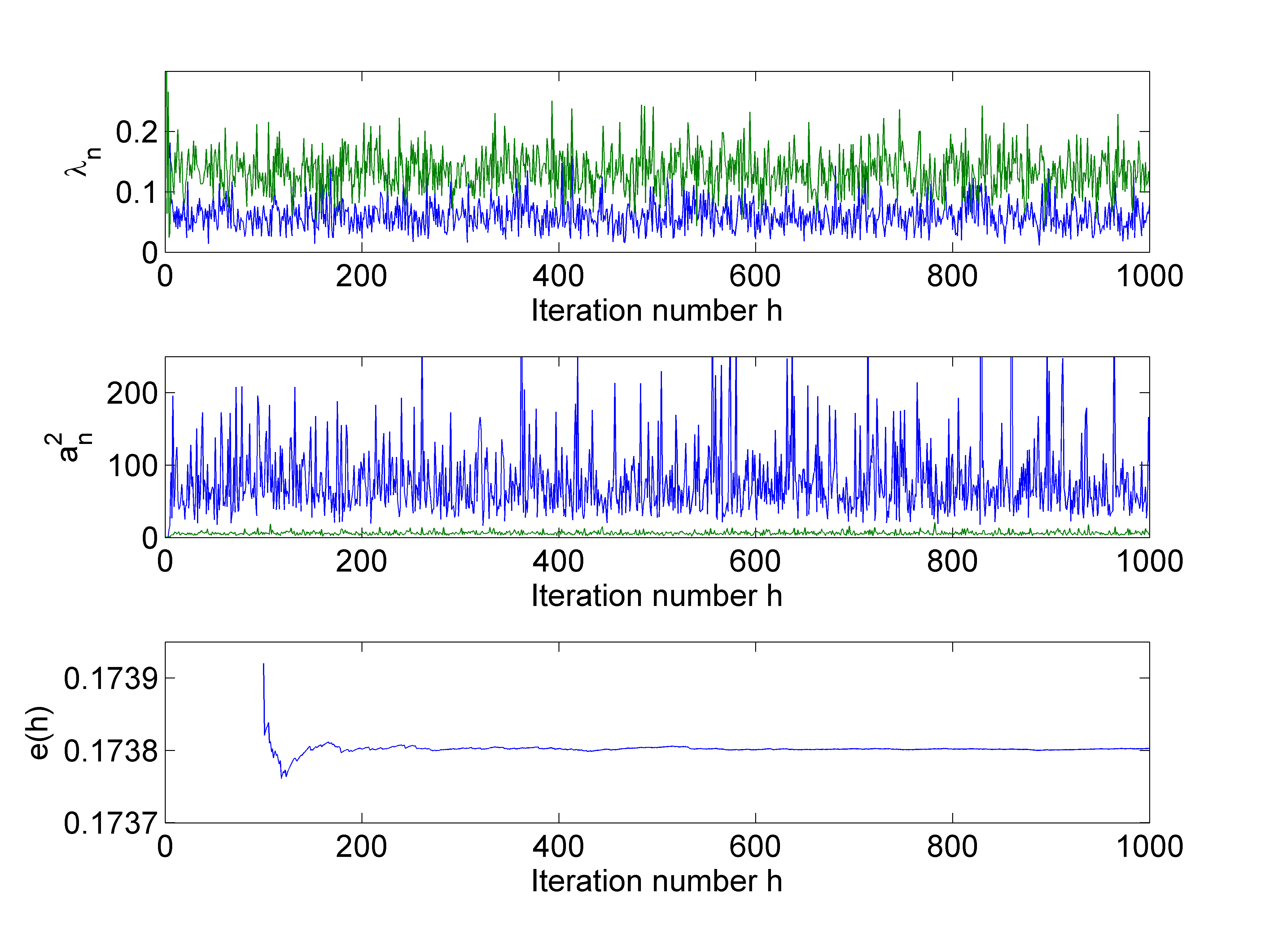}
  \caption{Top (resp. middle): source parameters $\souproba{\nosou}$ (resp. $\souvar{\nosou}$) generated by the proposed Gibbs sampler
    for the $1$st (blue) and $2$nd (green) sources.
    Bottom: error reconstruction as a function of iteration number.}
  \label{fig:result_convergence_SNR=15dB}
\end{figure}

The $\nbobs=100$ observation vectors are corrupted by an additive
white Gaussian noise with variance $\noisevar = 1.3 \times 10^{-3}$,
corresponding to a signal-to-noise ratio (SNR)
$\SNR_{\mathrm{dB}}=15\mathrm{dB}$ where
\begin{equation}
  \textrm{SNR}_{\mathrm{dB}} = 10 \log_{10}\left(\frac{\sum_{\noobs=1}^{\nbobs}
\norm{\MATaxe\Vsou{\noobs}}^2}{\dimobs\nbobs\noisevar}\right).
\end{equation}
The proposed Gibbs algorithm is applied on these noisy observations
with $N_{\mathrm{MC}}=1000$ iterations including
$N_{\mathrm{bi}}=100$ burn-in iterations. Note that these numbers of
iterations have been chosen to ensure convergence of the Markov
chains. More precisely, as a first convergence assessment, the
outputs of the Markov chains have been monitored for different
parameters of interest. As examples, the source parameters
$\souproba{\nosou}$ and $\souvar{\nosou}$ ($\nosou=1,2$) generated
by the proposed BOCA algorithm have been depicted as functions of
the iteration number in Fig.~\ref{fig:result_convergence_SNR=15dB}
(top and middle, respectively). Note that the generated values
converge towards the actual values of the corresponding parameters
after very few iterations. As an additional convergence criterion,
the reconstruction error
\begin{equation}
  e(h) = \sqrt{\sum_{\noobs=1}^{\nbobs}\norm{
\Vobs{\noobs}-\hat\MATaxe^{(h)}\hat\bfs^{(h)}(\noobs)}^2}
\end{equation}
has been evaluated as a function of the iteration number $h$
($h=N_{\textrm{bi}}+1,\ldots,N_{\textrm{MC}}$) where
$\hat\MATaxe^{(h)}$ and $\hat\bfs^{(h)}(\noobs)$ are the MMSE
estimates of $\MATaxe$ and $\Vsou{\noobs}$ computed following
\eqref{eq:MMSE_estimate} after $h$ iterations, respectively. The
results depicted in Fig.~\ref{fig:result_convergence_SNR=15dB}
(bottom) show that $N_{\mathrm{MC}}=1000$ iterations are sufficient
to ensure a small reconstruction error. The computation time
required by $1000$ MCMC iterations is $17$s for an unoptimized
MATLAB 2007b 32-bit implementation on a 2.2-GHz Intel Core 2. Of
course, for more challenging problems, more MCMC iterations may be
required.

The obtained joint MAP estimates of the sources and mixing matrices
are represented in Fig.~\ref{fig:result_sources_SNR=15dB} (stars,
blue) and Fig. \ref{fig:result_psi_SNR=15dB} (blue), respectively.
These results show that the proposed BOCA algorithm allows one to
estimate the sources and basis vectors for this simple example. Note
that the first active component in the second source signal has not
been detected due to its very low amplitude.

\begin{figure}[h!]
  \centering
  \includegraphics[width=\figwidth]{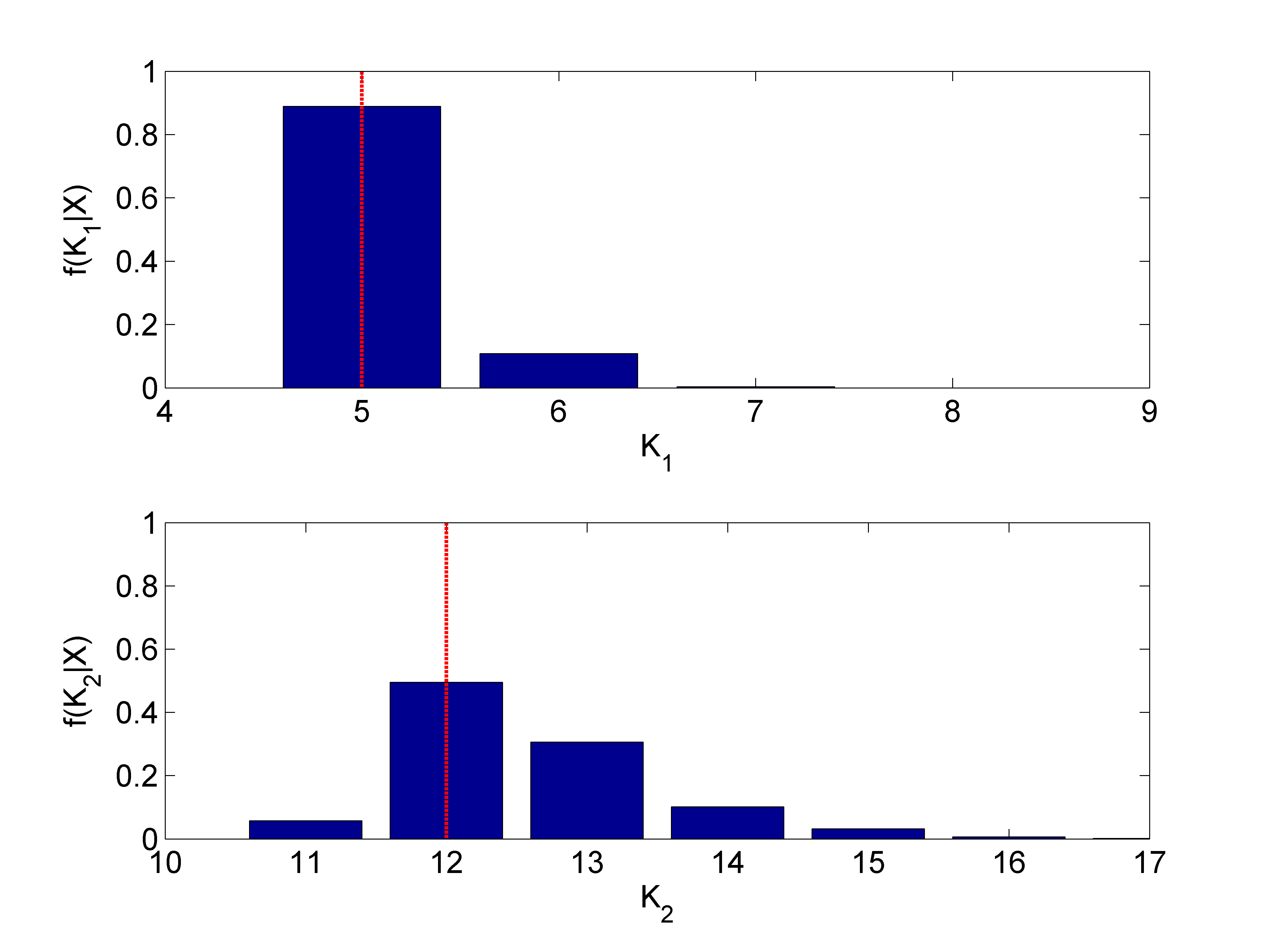}
  \caption{Estimated histograms of numbers of active components in the sources. The actual numbers appear in red dotted line.}
  \label{fig:result_K_SNR=15dB}
\end{figure}

The proposed algorithm implicitly generates binary variables
$\indic{\nosou}{\noobs}$ that indicate the presence/absence of
non-zero source components (see paragraph
\ref{subsec:sample_sources}). Thus, these indicators can be used to
compute interesting statistics regarding the probability of having
active components. As an example, the number $K_\nosou$
($\nosou=1,\ldots,\nbsou$) of active components in a given source
$\Vsou{\nosou}$ can be estimated by
\begin{equation}
  K_\nosou = \sum_{\noobs=1}^{\nbobs} \indic{\nosou}{\noobs}.
\end{equation}
The posterior probabilities of $\hat K_1$ and $\hat K_2$ estimated
by the proposed method for the considered synthetic example are
represented as histograms in Fig.~\ref{fig:result_K_SNR=15dB}. The
actual numbers of active components $K_1=5$ and $K_2=12$ are
represented by red dotted lines in these figures.

\begin{figure}[h!]
  \centering
  \includegraphics[width=\figwidth]{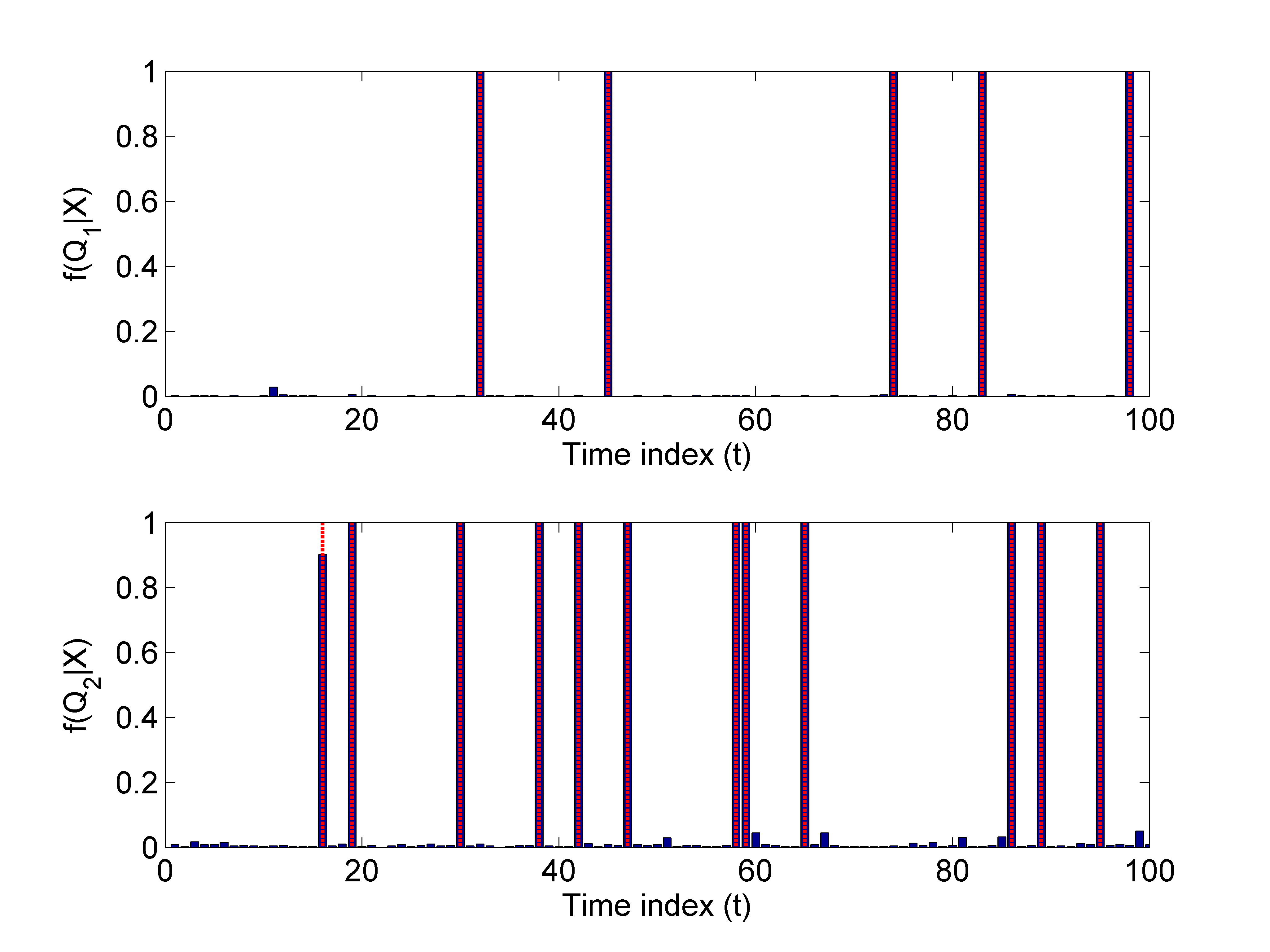}
  \caption{Posterior probabilities of having active components. The actual active non-zero components appear in red dotted line.}
  \label{fig:result_Q_SNR=15dB}
\end{figure}

The binary variables $\indic{\nosou}{\noobs}$ have Bernoulli
distributions. Therefore, the MMSE estimator of
$\indic{\nosou}{\noobs}$ provides the posterior probability of
$\sou{\nosou}{\noobs}$ to be active. Following
\eqref{eq:MMSE_estimate}, the MMSE estimates of the indicators
$\indic{\nosou}{\noobs}$ are computed and represented in
Fig.~\ref{fig:result_Q_SNR=15dB}. These posterior probabilities are
in good agreement with the actual positions of the active
components, represented by red dotted lines in these figures. Note
that these probabilities allow one to locate the first active
component in the second source signal that has been previously
omitted by the MAP estimator.

\subsection{Performance comparison}
\label{subsec:performance_comparison} 
\begin{figure}[h!]
  \centering
  \includegraphics[width=\figwidth]{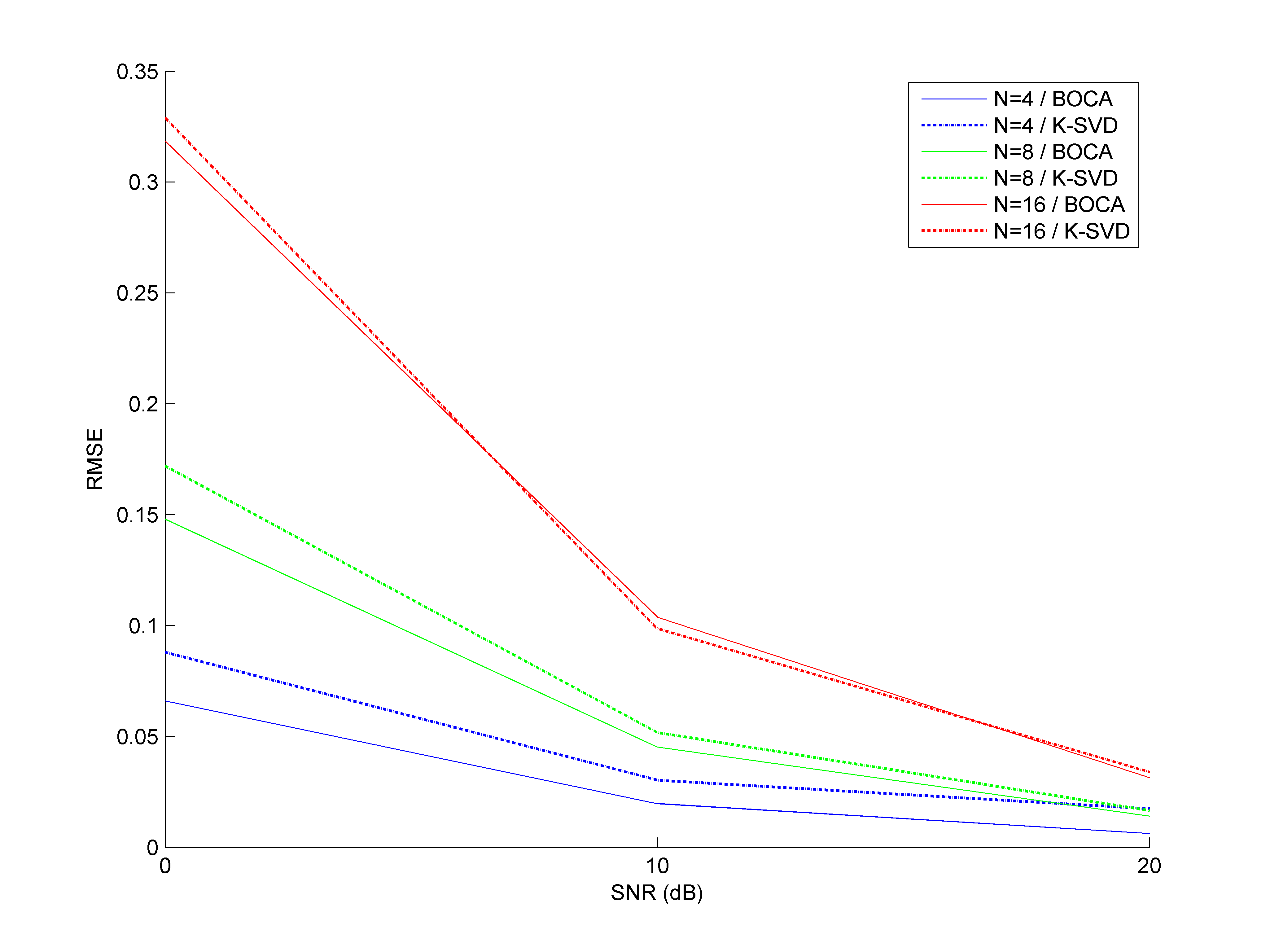}
  \caption{RMSEs as functions of the noise level $\textrm{SNR}$ for different values of the number of sources $\nbsou$.}
  \label{fig:result_MC_RMSE}
\end{figure}

\begin{figure}[h!]
  \centering
  \includegraphics[width=\figwidth]{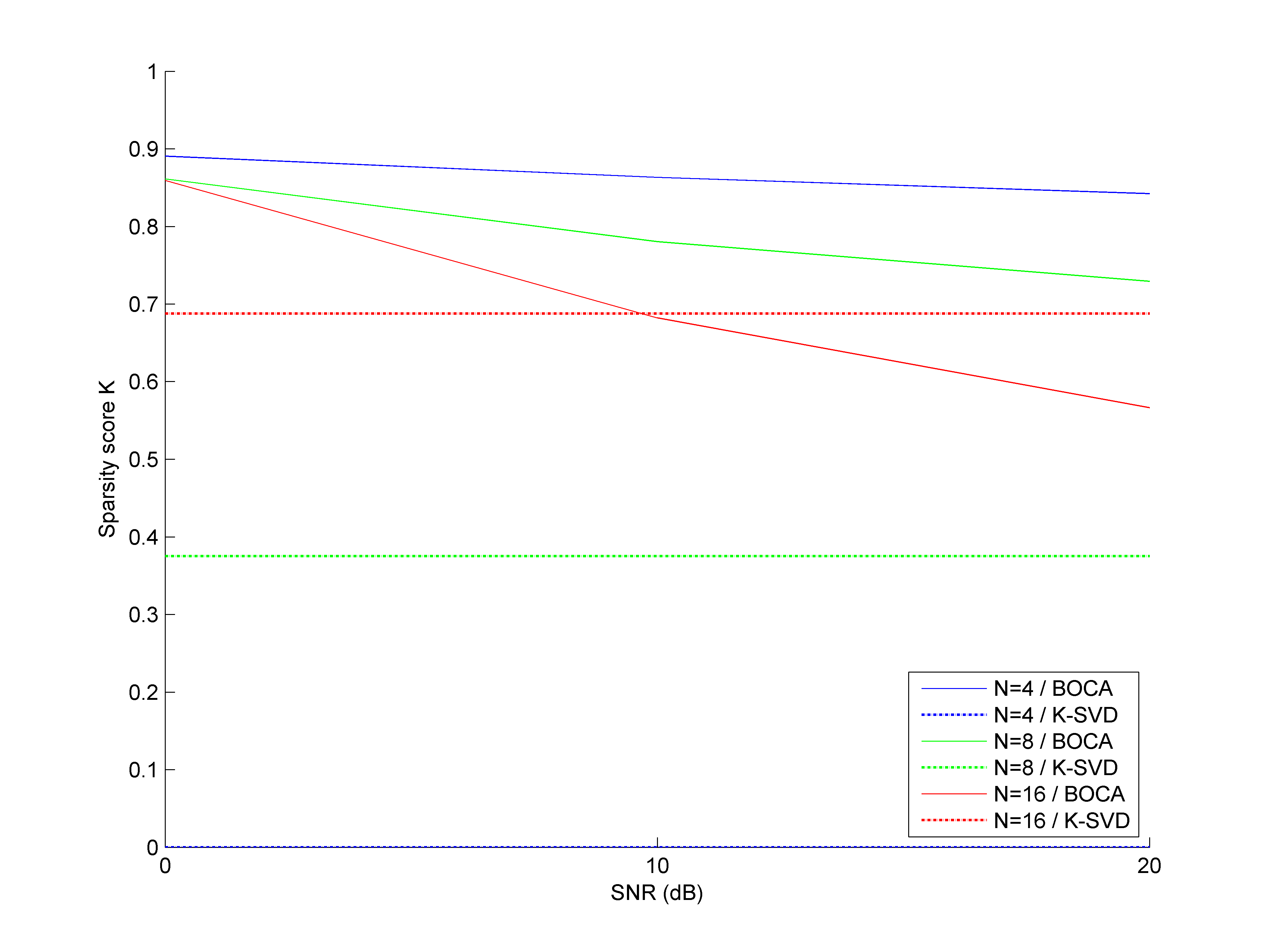}
  \caption{Sparsity as function of the noise level $\textrm{SNR}$ for different values of the number of sources $\nbsou$.}
  \label{fig:result_MC_score_K}
\end{figure}

We propose here to compare the
proposed algorithm with an up-to-date dictionary learning technique
referred to as K-SVD algorithm \cite{Aharon2006}. This algorithm has
been widely applied for various signal and image processing problems
and has demonstrated promising results (see for example
\cite{Bertin2007}, \cite{Bryt2008}, and \cite{Mairal2008}). For
fixed signal dimensions $\dimobs=128$ and $\nbobs=256$, an $\dimobs
\times \nbobs$ sparse matrix $\MATsou$ is randomly drawn according
to the prior in \eqref{eq:prior_sources} with
$\souproba{1}=\ldots=\souproba{\nbsou}=0.05$ and $\souvar{1}
=\ldots=\souvar{\nbsou}= 10$. Then, an $\dimobs \times \nbsou$
random orthogonal matrix $\MATaxe$ is selected as the first $\nbsou$
columns of the left orthogonal matrix provided by the
singular-value-decomposition (SVD) of an $\dimobs\times\dimobs$
matrix whose elements have been drawn according to a
$\calN\left(0,1\right)$ distribution. The number of dictionary atoms
$\nbsou$ is set to three different values: $\nbsou=4, 8$ and $16$.
The observations, computed following \eqref{eq:model_matrix}, are
corrupted by an additive Gaussian noise with SNR ranging from $0$dB
to $20$dB. The proposed Bayesian algorithm is applied on the
generated data with $N_{\textrm{MC}}=300$ Monte Carlo iterations and
$N_{\mathrm{bi}}=50$ burn-in iterations. The accuracy of the MAP
estimates of the source and mixing matrices
$\hat\MATsou_{\textrm{MAP}}$ and $\hat\MATaxe_{\textrm{MAP}}$ are
compared with the results provided by the K-SVD algorithm with a
total number of $80$ iterations (as in \cite{Aharon2006}). The
performance of these algorithms are expressed in terms of
reconstruction error and sparsity. More precisely, the root mean
square error (RMSE) between the actual noise-free data
$\MATaxe\MATsou$ and the estimated reconstructed data
$\hat\MATaxe\hat\MATsou$ is computed as
\begin{equation}
  \mathrm{RMSE} =
\sqrt{\frac{1}{\dimobs\nbobs}\sum_{\noobs=1}^{\nbobs}\norm{\MATaxe\Vsou{\noobs}-\hat\MATaxe\hat\bss\left(\noobs\right)}^2}.
\end{equation}
The sparsity of the estimated source vectors is measured using the
following score function $\tilde K$
\begin{equation}
\begin{split}
  \tilde K &= 1-\frac{1}{\nbsou\nbobs}\sum_{\nosou=1}^{\nbsou}
\norm{\hat\bss\left(\noobs\right)}_0\\
            &= 1-\frac{1}{\nbsou\nbobs}\sum_{\nosou=1}^{\nbsou}\sum_{\noobs=1}^{\nbobs}
\hat q_{\nosou}(\noobs).
\end{split}
\end{equation}
Note that $\tilde K \in [0,1]$ where $\tilde K = 1$ means that
$\hat\MATsou$ is the $\nbsou\times\nbobs$ matrix of zeros whereas
$\tilde K=0$ means that $\hat\MATsou$ is a matrix containing
$\nbsou\nbobs$ non-zero elements. The RMSEs and score functions
$\tilde K$, computed over $100$ Monte Carlo trials, are depicted in
Fig.'s \ref{fig:result_MC_RMSE} and \ref{fig:result_MC_score_K} as
functions of the noise level $\mathrm{SNR}$ and for different values
of the number of sources $\nbsou$.

Fig.~\ref{fig:result_MC_RMSE} indicates that the Bayesian orthogonal
component analysis (BOCA) generally outperforms the K-SVD algorithm
in term of reconstruction RMSE. In other words, the proposed
strategy provides a combination of source and mixing matrices
$\hat\MATsou$ and  $\hat\MATaxe$ that better fit the observed data
than the solution provided by K-SVD, especially at low SNR.
Moreover, Fig.~\ref{fig:result_MC_score_K} shows that when the
number of sources and the SNR are low, the sources estimated by BOCA
are much sparser than the sources identified by K-SVD. Note that the
sparsity level of the solutions provided by K-SVD is implicitly
fixed by the number of non-zero entries in the
orthogonal-matching-pursuit (OMP) sub-procedure, whereas the
introduced BOCA estimates this sparsity degree via an unsupervised
framework.

\section{Application to sparse coding with under-complete orthogonal dictionaries}
\label{sec:image_proc}

\begin{figure*}
  \centering
  \includegraphics[width=\textwidth]{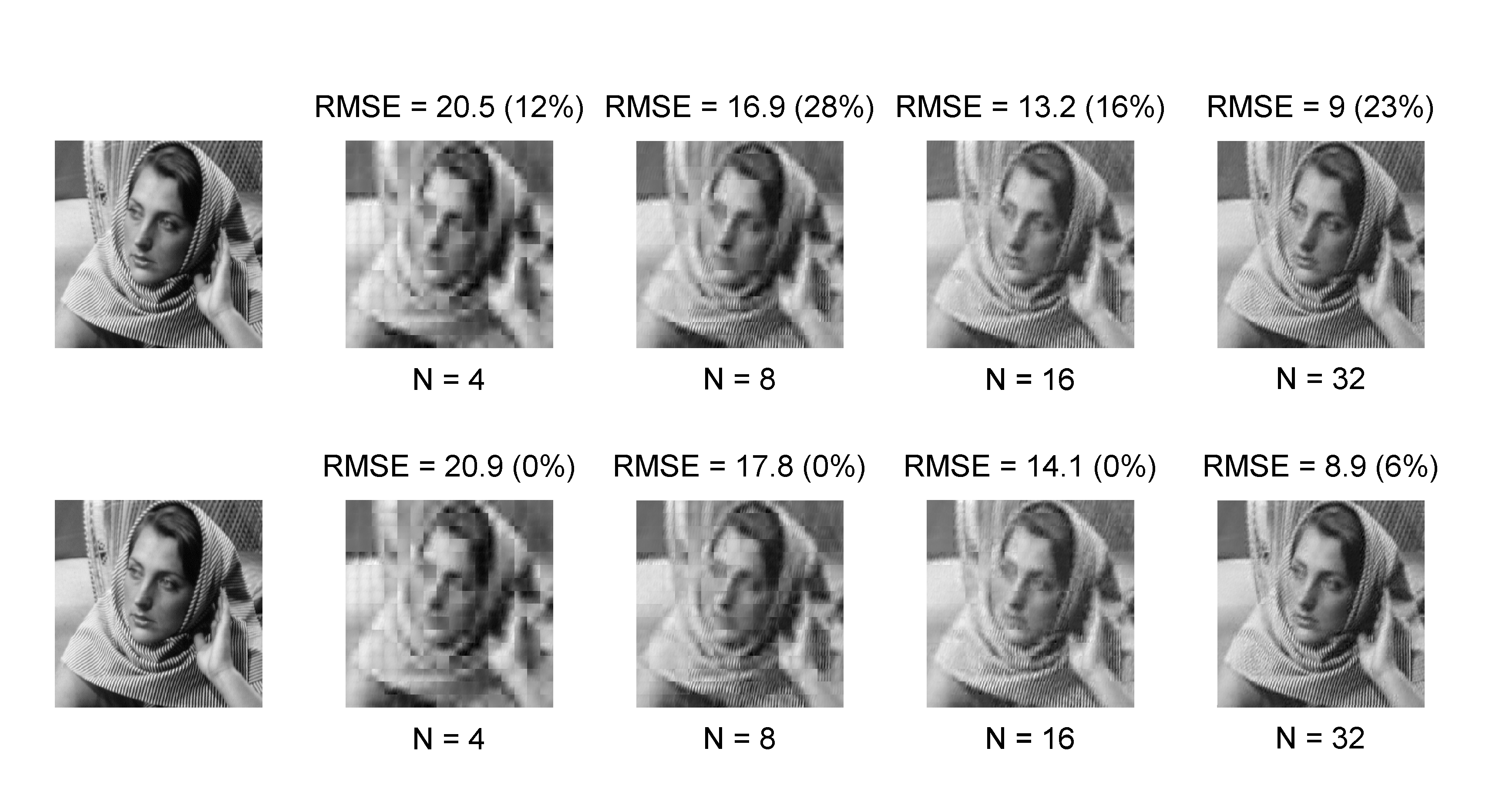}
  \caption{Sparse coding of the Barbara image obtained by BOCA ($1$st row) and K-SVD ($2$nd row) with different values of $\nbsou$.
    Corresponding RMSEs and sparsity levels (expressed as
    percentage) are indicated above each image.}
  \label{fig:result_paper_barbara}
\end{figure*}

In this section, we present the ability of the proposed procedure to
perform sparse coding with under-complete orthogonal dictionaries. A
fraction of the well-known Barbara natural image is analyzed by
BOCA. This $256 \times 256$-pixel image, depicted in
Fig.~\ref{fig:result_paper_barbara} (column $\# 1$), is decomposed
into $\nbobs = 16^2$ block patches of size $\dimobs = 16\times16$
pixels. The proposed Bayesian strategy and the K-SVD algorithm are
applied on these observations for different values of the number of
sources  $\nbsou$ (i.e., different numbers of dictionary atoms). The
images reconstructed by the algorithms after estimating the source
and mixing matrices are depicted in
Fig.~\ref{fig:result_paper_barbara} (column $\#2$--$5$) for
different values of $\nbsou$ ranging from $\nbsou=4$ to $\nbsou=32$.
As in paragraph~\ref{subsec:performance_comparison}, the estimation
performances of both algorithms are evaluated in terms of
reconstruction error (RMSE) and sparsity level ($\tilde K$ expressed
as a percentage). The RMSEs are reported for each reconstructed
image and the sparsity measures appear between brackets. These
results clearly indicate the reliability of BOCA for fitting the
observed data and its ability of identifying a sparse
representation.

\begin{figure}[h!]
  \centering
  \includegraphics[width=\figwidth]{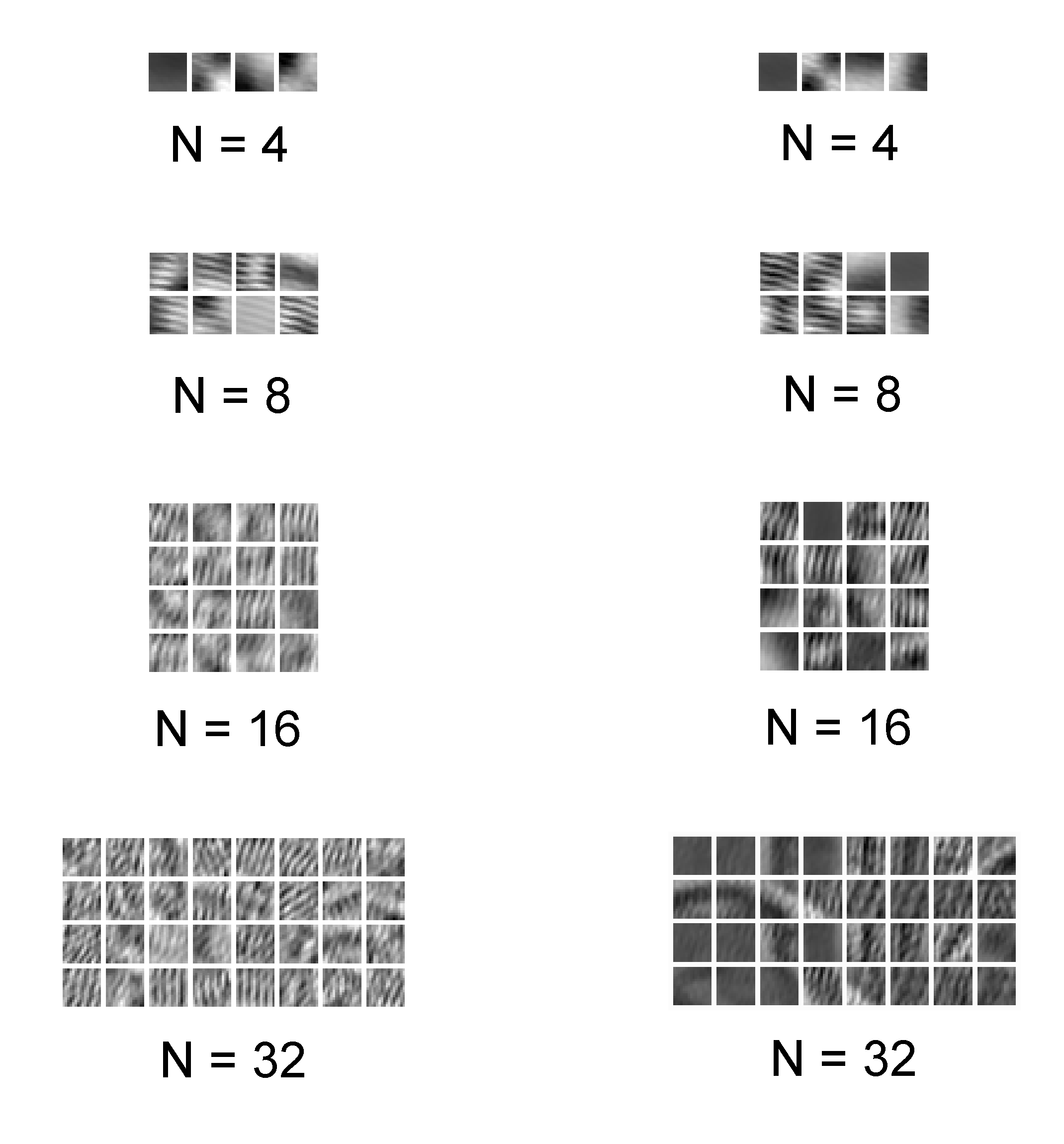}
  \caption{Estimated dictionary atoms by BOCA (left) and K-SVD (right) for different values of the number of dictionary atoms $\nbsou$.}
  \label{fig:result_paper_barbara_dictionary}
\end{figure}

The Bayesian algorithm generates a collection of mixing matrices
$\left\{\sample{\MATaxe}{h}\right\}_{h=1,\ldots,N_{\textrm{MC}}}$
that can be used to approximate the MAP estimator of $\MATaxe$
following \eqref{eq:MAP_estimate}. The MAP estimate of the
dictionary atoms (i.e., the mixing matrix), formatted as $\nbsou$
block patches of size $16\times 16$, are represented in
Fig.~\ref{fig:result_paper_barbara_dictionary} (left) for the
corresponding values of the number of sources $\nbsou$. As an
illustration, the dictionary atoms estimated by K-SVD have been also
depicted in Fig.~\ref{fig:result_paper_barbara_dictionary} (right).

\section{Conclusions and future work}
\label{sec:conclusion} We introduced in this paper a new Bayesian
algorithm for sparse representation with under-complete orthogonal
dictionary. This problem was formulated as a blind separation
problem of sparse sources mixed by an orthogonal matrix. The
proposed approach relied on appropriate  prior distributions for the
unknown model parameters. The sparse sources to be estimated were
modeled as Bernoulli-Gaussian processes. A uniform distribution on
the Stiefel manifold was elected as prior distribution for the
mixing matrix. The hyperparameters associated with this prior model
were estimated from the data in an unsupervised fully Bayesian
framework. A partially collapsed Gibbs sampler was studied to
generate samples distributed according to the joint posterior
distribution of the mixing matrix, source matrix, the noise variance
and the model hyperparameters. The Bayesian estimators of the
unknown model parameters were then approximated by using the
generated samples. The estimation performance of the proposed
algorithm was evaluated from simulations conducted on synthetic
data. A comparison with the K-SVD algorithm showed very promising
results in favor of the proposed Bayesian method. An application of
the proposed sparse coding technique for natural image processing
was also investigated.

Future works include the unsupervised estimation of the number of
sources $\nbsou$ (i.e., dimension of the subspace) using a
reversible-jump MCMC algorithm as in \cite{Hoff2007}. Extension of
the proposed linear decomposition model to union of orthogonal
dictionaries as in \cite{Gribonval2003} is currently under
investigation. Finally, it would be interesting to apply BOCA to
sparse coding in transformed domains for compression problems.

\section*{Acknowledgment}
The authors would like to thank Prof. Alfred O. Hero (University of
Michigan) for interesting suggestions related to this work. They are
also very grateful to G\'eraldine Morin and Marie Chabert
(University of Toulouse) for their valuable feedback regarding the
application of BOCA on natural images considered in this work.

\bibliographystyle{ieeetran}
\bibliography{biblio}

\end{document}